\documentclass{ar-1col}
\usepackage{url,color}
\usepackage[sort&compress,numbers]{natbib}

\jname{Xxxx. Xxx. Xxx. Xxx.}
\jvol{AA}
\jyear{YYYY}
\doi{10.1146/((please add article doi))}

\graphicspath{{figures/}}

\def\De{\mbox{De}}
\def\Bn{\mbox{Bn}}
\def\ch{\textcolor{black}}

\begin{document}

\markboth{S. E. Spagnolie and P. T. Underhill}{Swimming in Complex Fluids}

\title{
\LARGE Swimming in Complex Fluids}

\author{Saverio E. Spagnolie$^1$ and Patrick T. Underhill,$^2$
\affil{$^1$Department of Mathematics, and Department of Chemical and Biological Engineering, University of Wisconsin-Madison, Madison, WI, USA, 53706; email: spagnolie@math.wisc.edu}
\affil{$^2$Department of Chemical and Biological Engineering, Rensselaer Polytechnic Institute, Troy, NY, USA, 12180; email: underhill@rpi.edu}}

\begin{abstract}
We review the literature on swimming in complex fluids. A classification is proposed by comparing the length and time scales of a swimmer with those of nearby obstacles, interpreted broadly, extending from rigid or soft confining boundaries to molecules which confer the bulk fluid with complex stresses. A third dimension in the classification is the concentration of swimmers, which incorporates fluids whose complexity arises purely by the collective motion of swimming organisms. For each of the eight system classes which we identify we provide a background and describe modern research findings. While some classes have seen a great deal of attention for decades, others remain uncharted waters still open and awaiting exploration.
\end{abstract}


\maketitle


\section{INTRODUCTION}

Biological fluids are messy. They can become more or less viscous by stirring them. They can store and release elastic energy across a wide range of relaxation timescales, or they can be soft in one direction and hard in another. Strange as they may be, these are the environments that microorganisms must often navigate in order to survive. A medium can even express astounding bulk features due to the emergent activity of the microorganisms themselves. The scientific literature on swimming in complex fluids is equally expansive - the phrase itself is used to describe physical systems which may seem hardly related.

The physics of swimming has captured the imagination for millenia \cite{gmv17}, but advanced theories treating the locomotion of bodies through fluids would not appear until the dawn of mathematical fluid mechanics, and then the confluence of physics and biology in the 20th century. The study of microorganism locomotion at the micron scale, where viscous dissipation dominates inertia, has now been a tremendously active research area for the better part of a century \cite{Lighthill76,lp09,ewg15,Lauga16,gwssnplgka20}.

The earliest works produced a great understanding of swimming in idealized (``Newtonian'') fluids. The fluids through which microorganisms swim, however, are often endowed with highly nonlinear, non-Newtonian features when viewed at a continuum scale, including shear-dependent viscosity, elasticity, and other features which blur the lines between fluids and solids. A broader understanding of the role of complex fluids in biological systems, including and beyond the physics of locomotion, is achieved with each passing year \cite{Spagnolie15}. The importance of these complex fluid phenomena on self-propulsion depends on the length- and time-scales of the body motion when compared to those of the environment. Just as the Reynolds number\footnote{The Reynolds number, $Re=\rho U L /\mu$, with $\mu$ the fluid's viscosity, and $L$ and $U$ characteristic length and velocity scales of the swimming body, is orders of magnitude smaller than 1 for microorganism locomotion \cite{Purcell77,Childress81}.} measures the relative importance of inertial and viscous effects, so the Carreau, Weissenberg, Bingham, and Ericksen numbers measure the importance of shear-thinning, viscoelasticity, viscoplasticity, and anisotropy \cite{Larson99}. 

Some fluids are even complex {\it as a consequence} of microorganism activity if treating the suspension of swimmers as part of the material. Active suspensions or active matter can exhibit orientational order and elastic response, shear-dependent viscosity, and ``negative viscosity'' \cite{hrrs04, ip07, habk08, sa09,Saintillan18}. In addition to the microorganisms and suspending material, the presence of boundaries, whether internal to the fluid or confining it externally, can contribute to the emergence of the complex features noted above. For a broader look at the literature on confinement of active particles see the review by Bechinger et al.~\cite{bdlrvv16}.

Given the many interpretations and dimensions of ``swimming in complex fluids'' a classification system would seem to be of use. Such a system may help identify the most important features of a physical system and unify seemingly disparate effects. The purpose of this review is twofold. First we aim to outline such a classification for swimming in complex fluids. To identify distinct classes we will consider the relationship of the swimmer(s) to the constituents or obstacles that provide the fluid with complex features through their relative lengths and timescales, and consider the concentration of swimmers. The second goal of this paper is to provide a review of the literature for each of these classes, and to draw connections between them. While our classification is not all-encompassing, we hope that it will provide organizational clarity on the tremendously diverse field of physical phenomena and biological behavior.

\section{A CLASSIFICATION OF SWIMMING IN COMPLEX FLUIDS}
\noindent \textit{It was the misfortune of the {\it Proteus} and her crew to be pioneers into a realm that was literally unknown; surely a fantastic voyage if ever there was one.} \cite{akk66}
\vspace{.2cm}

\begin{figure}[htbp]
\begin{center}
\includegraphics[width=\textwidth]{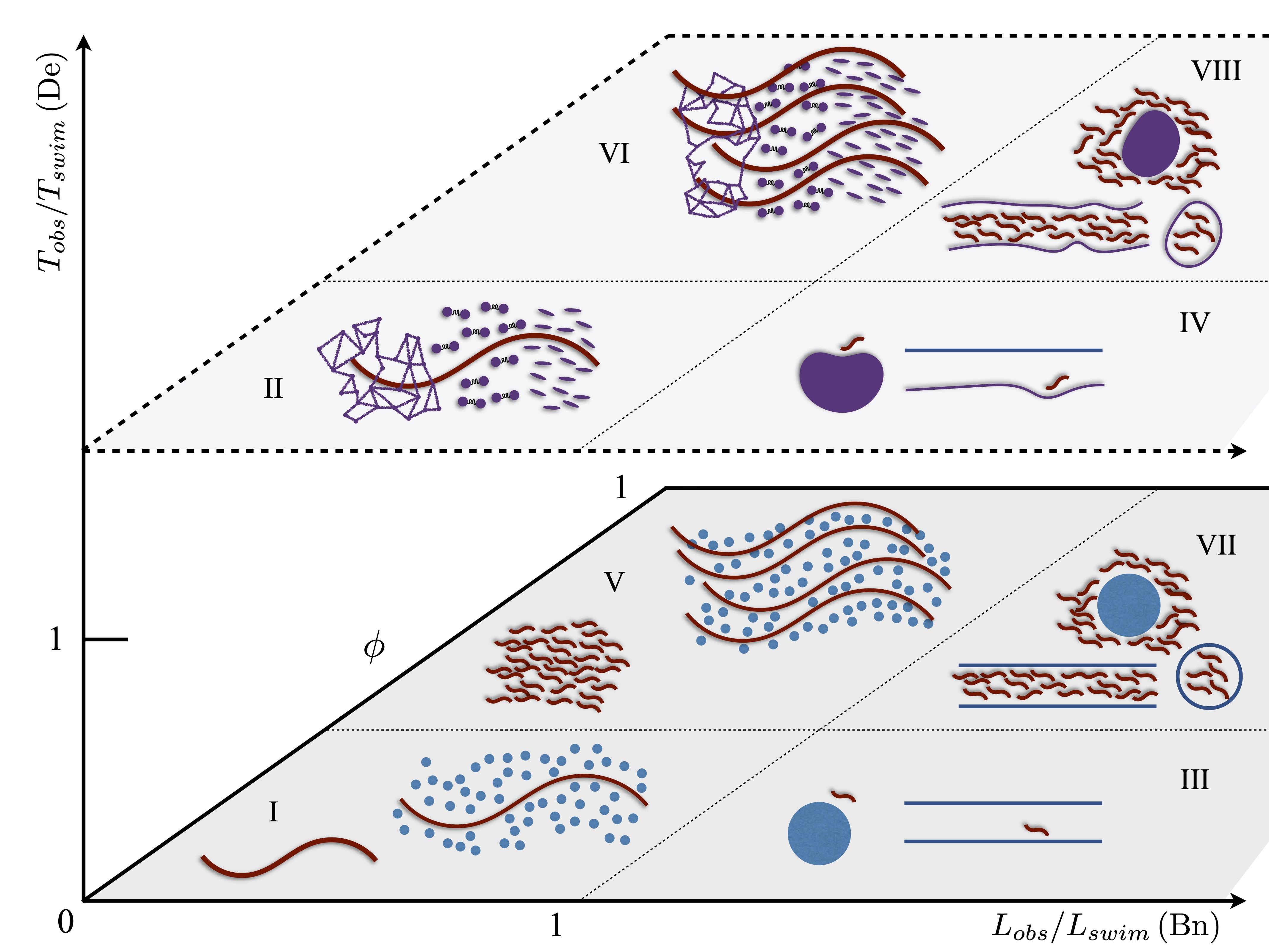}
\caption{A classification of swimming in complex fluids. The axes are the relative lengths of the obstacle(s) to the swimmer, $\Bn=L_{obs}/L_{swim}$, their relative timescales, $\De=T_{obs}/T_{swim}$, and the volume fraction of swimmers $\phi$. Red objects are active (swimming), blue objects are rigid, and purple objects are deformable. I/II:~a single swimmer in a suspension of small rigid/deformable obstacles; III/IV: a single swimmer near a large rigid/deformable obstacle; V/VI: many swimmers in a suspension of small rigid/deformable obstacles; VII/VIII: many swimmers near a large rigid/deformable obstacle.}
\label{fig: classification}
\end{center}
\end{figure}

With the aim of organizing the complex fluid phenomena which either affect or are caused by swimming microorganisms, we focus on the relationship between an individual swimmer and its environment. The environment is characterized by the distribution and behavior of ``obstacles'', broadly interpreted. There may be only one large obstacle, representing a rigid or deformable boundary. Or the obstacles might be much smaller than the swimming body, like polymers or long-chain proteins, which relax into their preferred configurations over an intrinsic set of timescales. When there are sufficiently many such obstacles they confer the fluid with bulk features like viscoelasticity and shear-dependent viscosity. The obstacles might move with a background fluid, might themselves compose the background fluid, or they might be nearly fixed in space, for example if the obstacle is large or if it is composed of a cross-linked network of fibers.

\textbf{Figure~\ref{fig: classification}} shows eight extremes of interest in our classification system, organized in a three-dimensional space. The vertical axis is the relaxation timescale of the obstacle(s) relative to a timescale associated with swimming motion, $T_{obs}/T_{swim}$, often called the Deborah number, $\De$ \cite{Reiner64}. Depending on the context and the physical processes that produce the timescales, this ratio might instead be referred to as a Weissenberg number, or possibly by other names, as will be discussed. The horizontal axis is the length scale of the obstacle(s) relative to a length associated with the swimming body, $L_{obs}/L_{swim}$, which we term the Benes number\footnote{In the fictional ``Fantastic Voyage'' by Jerome Bixby and Otto Klement, novelized by Isaac Asimov, Dr.~Jan Benes perfects a technology to manipulate an object's size \cite{akk66}. His body's various complex fluids go on to pose swimming and other challenges to the miniaturized crew injected into him to save his life.}, $\Bn$. Finally the third axis is the volume fraction of swimmers, $\phi$. The location of a system in this three-dimensional space influences the relative importance of different effects and the behaviors observed, and from a mathematical perspective the modeling approximations which are most appropriate. The ratio of length scales and concentrations determines whether the swimmers or obstacles appear as an effective continuum. The ratio of frequencies or time scales determines the extent to which the suspending medium's relaxation properties affect the dynamics. The concentration of swimmers determines the importance of interactions among groups of swimmers or active particles.

 
The classification is not all-encompassing, and there are physical systems which could lie in multiple domains or sit outside of it altogether. For instance, an additional time scale (and classification axis) not considered here is associated with the rate of viscous dissipation and determines the relative importance of inertia. Nevertheless we press on to discuss some of the fundamental features of the proposed organization. We begin with Type I systems, which at the extreme includes a single swimmer in a Newtonian fluid. 

\section*{$\mathcal{I}$. Small $\De$, small $\Bn$, small $\phi$: a single swimmer and small rigid obstacles}

In Type I systems, a single swimmer navigates an environment containing small obstacles which are effectively rigid on the timescale of swimming. At the extreme separation of length scales, far to the right in \textbf{Figure~\ref{fig: classification}}, the obstacles might even be the fluid constituents at the molecular scale; taking 10 microns as a characteristic swimmer size and comparing to the size of a single water molecule, $\Bn \approx 10^{-3}$. Type I therefore includes swimming in classical Newtonian fluids, a familiar territory from which to depart on our journey. 

The pioneering work on self-propulsion in viscous fluids appeared alongside improvements in microscope technology in the 1950s-1970s. Mathematical theories revealed many key features of swimming in Newtonian fluids at the microscale. They include instantaneously force- and torque-free dynamics, viscosity-independent swimming speeds for fixed gaits, kinematic reversibility (the ``Scallop theorem'' \cite{Purcell77}), and widespread use of slender filament drag anisotropy \cite{bw77,Childress81,lp09,Lauga16}. Mathematical models from this era which remain mainstays across the field include Taylor's infinite swimming sheet \cite{Taylor51}, Lighthill and Blake's spherical envelope squirmer \cite{Blake71} and rotating helical model flagella \cite{Lighthill76}.

\begin{figure}[htbp]
\centering
\includegraphics[width=0.5\textwidth]{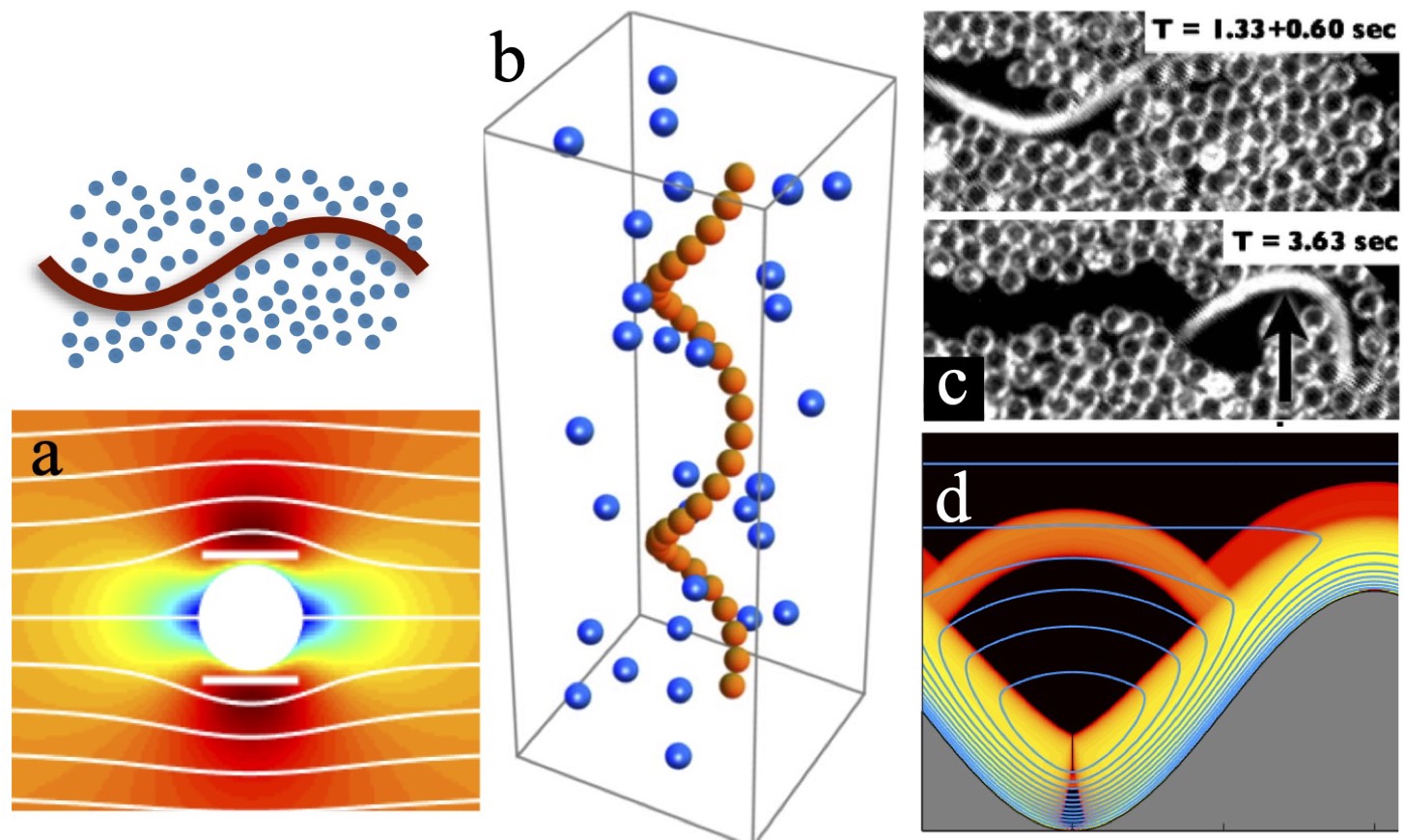}
\caption{Type I systems. (a) Velocity magnitude of flow past a cylinder and streamlines in a model porous (Brinkman) medium - the additional anisotropy results in enhanced swimming speeds \cite{lo16}. (b) A helical filament swims faster in a heterogeneous environment of free obstacles \cite{Leshansky09}. (c) {\it C. elegans} swims faster through a dense field of granular obstacles as it deforms the environment \cite{Jung10}. (d) Density map of the shear rate (on a logarithmic scale) near a swimming body in a viscoplastic medium at large Bingham number, showing local fluidization \cite{hb17}. All figures reproduced with permission.}
\label{fig: Figure_2}
\end{figure}

Moving away from the origin and this Newtonian extremity, the rigid obstacles in question increase in size relative to the swimmer size. If still small compared to the swimmer but large compared to the fluid's molecular constituents, they may still contribute features to the fluid which the swimmer experiences as a change in the bulk rheology. A porous medium, for example, which strongly affects the relationship between pressure and flow can in turn have an effect on a relatively large swimming body. \textbf{Figure~\ref{fig: Figure_2}a} shows the velocity magnitude of a flow past a rigid cylinder in a model porous (Brinkman) fluid, suggestive of a particularly strong drag anisotropy. Such an environment was found to enhance undulatory and helical self-propulsion \cite{lo16,hlo19}. But decreased speeds were observed in the same medium for soft, deformable swimmers \cite{lo16,hlo19} and for model tangential squirmers \cite{np18}.

As the obstacles become larger they have an even greater impact on swimming organisms. If they are fixed in space their hydrodynamic reach is roughly independent of size - each appears as a ``Stokeslet'' singularity in the fluid flow, rather than as a higher-order ``stresslet'' disturbance field \cite{kk13}. Changes in swimming are even predicted at low obstacle volume fractions. Leshansky showed that the presence of stationary obstacles in the fluid increases the swimming speed and efficiency of a helical filament with a fixed rotation rate or even with fixed power input (\textbf{Figure~\ref{fig: Figure_2}b}), essentially by pushing off of the surrounding obstacles (through the fluid)  \cite{Leshansky09,cltp20}. Experiments with {\it C. elegans} showed too an enhanced mobility in such environments due to an increase in tangential/normal drag anisotropy (\textbf{Figure~\ref{fig: Figure_2}c}) \cite{Jung10}. 

Greater obstacle volume fractions can result in additional complex features. The speed and efficiency of swimming in a viscoplastic fluid, which flows only beyond a critical stress, is characterized by the Bingham number, $\mbox{Bi}$, a characteristic ratio of the yield and viscous shear stresses. \textbf{Figure~\ref{fig: Figure_2}d} shows the density of fluid shear rate $\dot{\gamma}$ on a logarithmic scale near a swimming sheet at large Bingham number. Locomotion is critically dependent upon the fluidization of the environment local to the body surface and may involve the transport of plugged material regions \cite{hb17,Hewitt22}. 

Type I problems also include systems with a yet broader view of swimming, like the locomotion of sandfish lizards and snakes through dry granular media \cite{mdlg09,hg15}. So long as the relaxation time of the microstructure is small compared to the body's rate of motion, swimming through fluids with other unusual behavior may still appear as a Type I system. One such example is swimming in media with a fixed orientational order (e.g. a transversely isotropic fluid, \cite{ksp15,cds17,sp17}). The local order often has a longer relaxation timescale, however, which takes us in the direction of Type II systems.

\section*{$\mathcal{II}$. Large $\De$, small $\Bn$, small $\phi$: a single swimmer in a suspension of small deformable obstacles}

\noindent \textit{The antibodies lined up side by side, their spaghetti strand projections entangling.} \cite{akk66}
\vspace{.2cm}

In Type II systems the obstacles express their own relaxation features in a more significant way. They might be long-chain polymers, elastic networks or gels, or molecules which, though individually rigid, slowly relax together into order or to a disordered equilibrium state. Mammalian spermatozoa encounter several such complex fluids, including glycoprotein-based cervical mucus, mucosal epithelium inside the fallopian tubes, and actin-based viscoelastic gel outside the ovum \cite{sp06,fd06}. The relaxation may also be diffusive in nature; a dilute suspension of rigid rods may become more aligned by a flow, only returning to a isotropic orientational distribution on a timescale set by the fluid viscosity and temperature. 

\subsection*{Shear-dependent viscosity} 
Shear-dependent viscosity, shear-thinning in particular, appears often when a solvent is host to a suspension of small immersed obstacles or highly deformable polymers. A suspension of elongated particles tends to align with a background shear flow, reducing the bulk viscous resistance. The Carreau number $\mbox{Cu}=\lambda \dot{\gamma}$ is a dimensionless rate of the timescale $\lambda$ of the environment's relaxation process and the timescale associated with the local shear rate, $\dot{\gamma}^{-1}$ \cite{ms15}. Mucus, for instance, is significantly shear-thinning - with a relaxation time $\lambda$ on the order of 1000 s, its viscosity begins to drop at $\dot{\gamma}\approx 10^{-3}$ s$^{-1}$, then decreases by four orders of magnitude until $\dot{\gamma}\approx 10^{2}$s$^{-1}$ \cite{vl13,vf15}. In a swimming system the relevant shear rate is often that generated by the motion of a swimming body, and the Carreau number is precisely the comparison of timescales $T_{obs}/T_{swim}$ in \textbf{Figure~\ref{fig: classification}} (i.e. it could as well be called the Deborah number if the flow is oscillatory, or in a constantly sheared fluid the Weissenberg number, introduced below). 

\begin{figure}[htbp]
\centering
\includegraphics[width=0.8\textwidth]{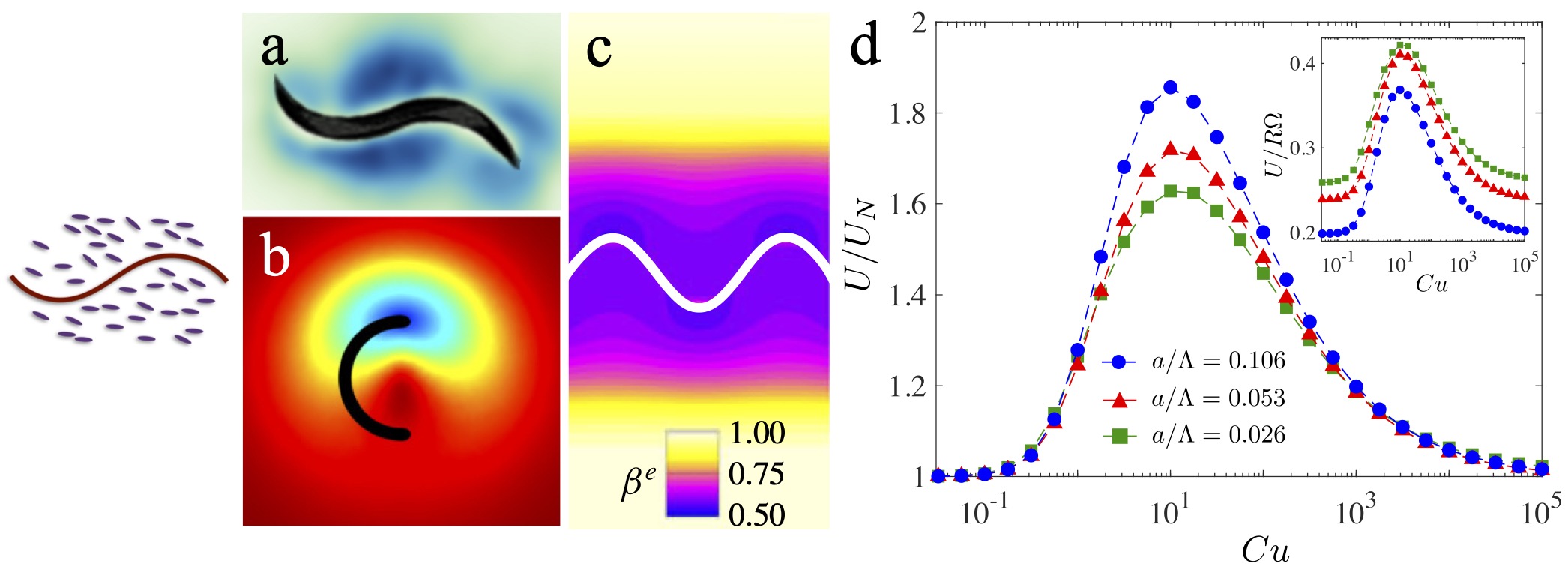}
\caption{Type II systems: shear-dependent viscosity. (a) Velocity magnitude near a swimming {\it C. elegans} in a shear-thinning fluid \cite{gka14}. (b) Viscosity field due to the prescribed motion of a helical body in a shear-thinning fluid, resulting in swimming enhancement \cite{dldp20}. (c) Similar viscosity stratification near a swimming sheet, also resulting in increased swimming speeds, with $\mbox{Cu}$ the Carreau number and $\beta^e$ the viscosity relative to the zero shear rate viscosity \cite{la15}. (d) A Goldilocks zone of Carreau numbers $\mbox{Cu}$ for confinement-like swimming enhancement of helical bodies for a few filament aspect ratios \cite{dldp20}. All figures reproduced with permission.}
\label{fig: Figure_3}
\end{figure}

Depending on the fluid, swimming in such an environment could instead be classified as a Type I system, as the obstacles themselves need not have an intrinsic deformability or timescale to confer shear-dependent bulk properties. Even in the relatively simple environment composed of a solvent and a suspension of shear aligning rigid rods, Brownian fluctuations introduce a relaxation timescale at the ensemble level. Fluid deformability might in this case be interpreted as the malleability of the distribution of obstacle orientations. A ratio of diffusive relaxation and flow timescales in this setting is generally called a Peclet number, $\mbox{Pe}$.

Experiments with the workhorse of undulatory locomotion, the nematode {\it C. elegans}, showed no substantial change in either the body kinematics or swimming speed in a shear-thinning Xanthan gum polymeric solution \cite{gka14} (\textbf{Figure~\ref{fig: Figure_3}a}). In a polystyrene colloidal suspension, however, the stroke form changed and swimming speeds increased by up to 12\%, while swimming efficiency continued to increase even once the swimming speed saturated \cite{pksw16}. In this second effort it was suggested that the effects were more apparent due to a larger viscosity contrast near the shear rates relevant to the swimming motion. 

Helical swimmers in a shear-thinning fluid showed even greater speed increases of up to 50\% in experiments \cite{gglz17} and beyond in simulations (\textbf{Figure~\ref{fig: Figure_3}b}) \cite{dldp20}. Similar results were found using model swimming sheets \cite{la15}. \textbf{Figure~\ref{fig: Figure_3}c} shows the viscosity relative to the zero shear rate viscosity at large Carreau number due to the undulatory motion of a swimming sheet with a prescribed motion, showing a very large region of diminished viscosity. Rather than being the result of surface-localized viscosity reduction, it has been suggested that the speed enhancement is due to a confinement-like effect owing to viscosity stratification. The more viscous regions appear as confining walls, apparent in \textbf{Figure~\ref{fig: Figure_3}b,c}, revealing a connection between Type II systems and Type III systems \cite{la15,ml15,gglz17}. The large amplitude body motions used in these numerical studies were also important - the effects of shear-dependent viscosity on the speeds of undulatory model swimmers enters only at fourth-order in the dimensionless amplitude \cite{vl13}.

The swimming speed relative to the Newtonian swimming speed for helical swimming in a shear-thinning fluid is shown in \textbf{Figure~\ref{fig: Figure_3}d} as a function of the Carreau number. For both small and large Carreau numbers the environment has a nearly constant viscosity everywhere and the fluid acts as an effective Newtonian fluid. Since the swimming speed of a body in a Newtonian fluid is independent of viscosity for fixed kinematics, the Newtonian swimming speed is recovered at these extremes. In the ``Goldilocks zone'' of $\mbox{Cu}$ of order 1-10, however, when the relaxation rate is comparable to the rate of body motion, the viscosity change has a more detailed spatial structure and swimming enhancement is observed. Theoretical investigations using the model tangential squirmer, meanwhile, revealed a speed {\it reduction}, in this range of Carreau numbers \cite{dzep15}. The details of how the boundary conditions and the resulting viscosity stratification affects swimming is thus not always easy to guess.

\subsection*{Viscoelasticity} 

\begin{figure}[htbp]
\begin{center}
\includegraphics[width=\textwidth]{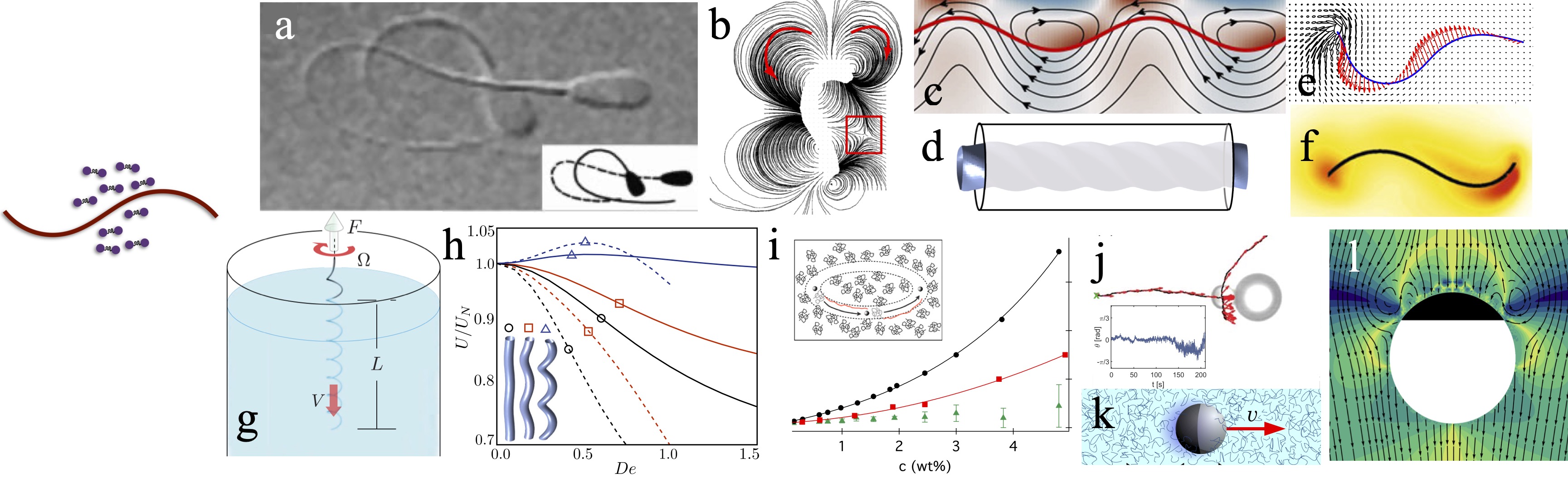}
\caption{Type II systems: viscoelastic fluids. (a) Spermatozoa become hyperactivated and swim faster 
\cite{hs01}. (b) The motion of {\it C. elegans} creates stagnation points in the flow and swims more slowly \cite{sa11}. (c) Streamlines generated by a swimming sheet near a boundary in a viscoelastic fluid \cite{im17}. (d) Swimming and pumping are modified through the mode-dependent complex viscosity, even in a confined domain \cite{ls15}. (e) Focusing of polymeric stress near the tips of a swimming sheet of fixed gait \cite{tfs10}; and (f) with a stress-dependent gait \cite{tg14}. (g) Schematic of an experiment showing enhanced propulsion of helical filaments in viscoelastic fluids \cite{lpb11}. (h) Scaled swimming speed across Deborah numbers for three different helical amplitudes, by helical waves (solid lines) and rigid rotation (dashed lines) from simulations \cite{slp13}. (i) Viscosity of a polymeric fluid vs. polymer concentration using a rheometer (circles), microrheology with a large bead (squares) and a flagellum diameter sized bead (triangles), suggesting strong shear-thinning at that scale (inset) \cite{msrwmp14}. Tangential squirmers in viscoelastic fluids: (j) with geometric confinement \cite{ngb19}; (k) showing increased rotational diffusivity \cite{gbb16}; and (l) driven by autophoresis \cite{ndhe17}. All figures reproduced with permission.}
\label{fig: Figure_4}
\end{center}
\end{figure}

Shear-dependent viscosity is among the simplest non-Newtonian bulk fluid features to consider. Short-range interactions among constituent molecules, or intrinsic relaxation times, introduce more dynamic bulk properties. How do such bulk features affect swimming organisms? Unlike in Newtonian fluids energy is temporarily bound into the fluid, which can either aid or resist self-propulsion. Extra fluid stresses can even change the shape of a swimming body or its motility organelles. Flagellar beating by spermatozoa becomes ``hyperactivated'' in viscoelastic fluids (\textbf{Figure~\ref{fig: Figure_4}a}) improving their mobility \cite{sd92}. But viscoelastic environments can reduce swimming speeds as well, as observed in experiments using {\it C. elegans} (\textbf{Figure~\ref{fig: Figure_4}b}) \cite{sa11}; see also the review by Sznitman and Arratia \cite{sa15}. 

Such fluids are commonly characterized by the Weissenberg number, $\mbox{Wi} = \lambda \dot{\gamma}$, yet again a product of a relaxation time $\lambda$ (here an elastic relaxation) and a shear rate $\dot{\gamma}$. In swimming problems, the shear rate is often oscillatory due to the periodic motion of motility organelles like flagella and cilia. Accordingly the Deborah number $\mbox{De}= \lambda \omega$ is more commonly used, where $\omega$ is the frequency of oscillation. The subtle distinction between the two depends on yet another length comparison. The magnitude of the shear rate in an oscillatory setting is $\dot{\gamma}=A\omega/L$, with $A$ the oscillation amplitude and $L$ the length scale of the oscillating body, thus $\mbox{Wi}/\mbox{De} = A/L$ which can be small or large in the same fluid. Many more details about Type II systems in particular can be found in recent reviews of swimming in viscoelastic fluids by Elfring and Lauga \cite{el15} and Li et al. \cite{lla21}, and of active colloids in complex fluids by Patteson et al.~\cite{pga16}. 

Viscoelastic model fluids like the Stokes/Oldroyd-B equations \cite{ms15} have been used to show that speeds should decrease monotonically with the Deborah number for infinite swimming sheets \cite{Lauga07,bcp10} and helical bodies \cite{fpw07,fwp09,ls15} upon the steady propagation of small amplitude waves. For small amplitude perturbations to the fluid, frequency space provides a natural decomposition of viscous and elastic effects, and explains why the same Deborah number dependence arises for both undulatory and helical motion \cite{el15,eg16}, even in confined systems (\textbf{Figure~\ref{fig: Figure_4}c,d}) \cite{ls15,im17}. Mobility enhancement in a purely viscoelastic fluid thus requires large amplitude body motions, finite length, or asymmetric beating, which have been explored numerically with fixed kinematics \cite{tfs10}, deformable bodies \cite{bcp10,tg14,rl14,bgs19}, and mixed wavespeeds \cite{rl15}. \textbf{Figure~\ref{fig: Figure_4}e,f} show the added polymeric stress near the tips of finite swimming sheets for fixed kinematics and for a deformable swimmer. These additional stresses can either aid or restrict propulsion depending on the nature of the stroke \cite{tfs10,tg14}. A spatially increasing wave amplitude introduces a flapping-like motion which would not generate a net thrust in a Newtonian fluid owing to kinematic reversiblity, but does in a viscoelastic fluid, leading to swimming enhancement \cite{pnl10}. Near walls, the fluid pinned between the swimmer and a surface can experience large local strains and stresses which in turn can have an outsized effect \cite{bcp10,lka14}. \ch{Similarly, such large strains near the body might be created by swirling flows, either by the body or by rotating flagella; the associated hoop stresses so generated can also provide a boost to swimming speeds \cite{bphs20,bs21}.}

Pathogen mobility in complex fluids is another area of interest for unsubtle reasons. Helical {\it Leptospira} and {\it B. burgdorferi} cells, the root cause of Weil's and Lyme diseases, respectively, swim faster in viscoelastic environments \cite{bt79,ks90}. Experiments using a Boger fluid \cite{lpb11} and numerical solution of the Stokes/Oldroyd-B equations \cite{slp13} recovered such enhancements at large helical amplitudes (\textbf{Figure~\ref{fig: Figure_4}g}). \textbf{Figure~\ref{fig: Figure_4}h} shows the swimming speed relative to the Newtonian swimming speed for helical bodies of different helical amplitudes across a range of Deborah numbers, with values over one achieved only beyond a critical large amplitude for both helical waves (solid lines) and rigid body rotation (dashed lines) \cite{slp13}. 

But relating experiments to simulations and theory can be complicated by the fact that shear-thinning and viscoelasticity often occur together in real fluids and can depend on distinct features acting simultaneously on multiple-scales \cite{msrwmp14, ml15}. \textbf{Figure~\ref{fig: Figure_4}i} shows the viscosity of a polymeric fluid as a function of polymer concentration using a rheometer (circles), microrheology with a 980 nm bead (squares) and microrheology using a small bead closer to the size of a flagellar cross-section (40 nm, triangles) showing no appreciable dependence on polymer concentration. On the scale of the flagellum, then, the resistance is merely due to the solvent viscosity, which was proposed as being due to severe shear-thinning in the region near the flagellum (inset) \cite{msrwmp14,zla18}. The fixed geometry of model tangential squirmers offers additional theoretical insight on the role of viscoelasticity, from the dependence on stroke asymmetry and geometric confinement (\textbf{Figure~\ref{fig: Figure_4}j}) \cite{lka14,ngb19}, to changes in rotational diffusivity and strong dependence on external forcing (\textbf{Figure~\ref{fig: Figure_4}k}) \cite{gbb16}, and how autophoretic motion couples to the complex generated flow (\textbf{Figure~\ref{fig: Figure_4}l})  \cite{ndhe17}. \ch{Novel methods and devices for swimming body rheometry continue to be developed, with an particular focus on future medical applications \cite{pzbl12,wcmpg20,kbeps22}. We refer the reader to the particularly germane review by Venugopalan et al. \cite{vepgw20}.}

\subsection*{Networks and gels} 

\begin{figure}[htbp]
\begin{center}
\includegraphics[width=\textwidth]{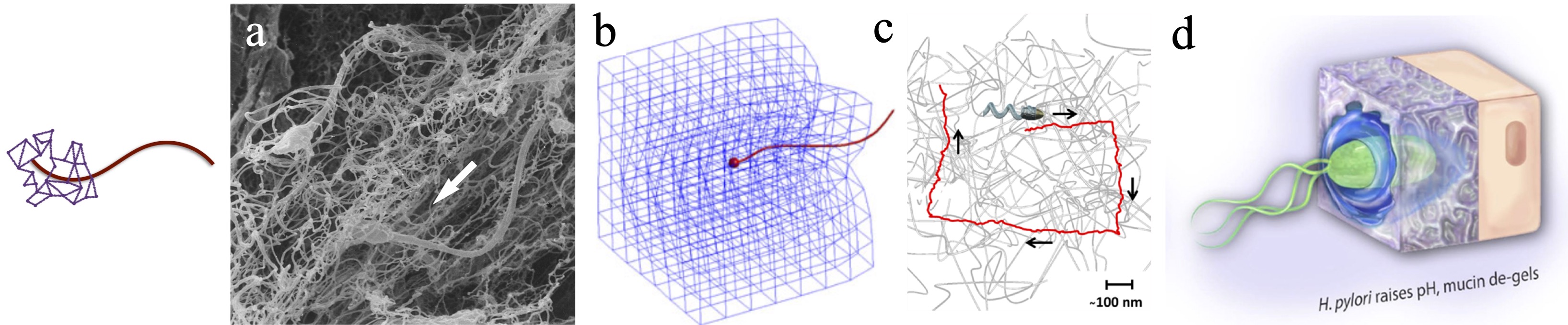}
\caption{Type II systems: networks and gels. (a) Spermatozoa swim through the anisotropic glycoproteic meshwork in cervical mucus \cite{Chretien03}. (b) Enhanced swimming of an undulating body through a stiff elastic network \cite{wlbfc16}. (c) A magnetically-driven synthetic swimmer probes a gel \cite{smgmmlf14}. (d) {\it H. pylori} chemically alters the rheology of its environment, improving its mobility \cite{ctakgkemse09}. All figures reproduced with permission.}
\label{fig: Figure_5}
\end{center}
\end{figure}

Rather than moving freely, the obstacles may instead be connected to each other. In this case the relevant relaxation timescale(s) may be that of a porous or cross-linked network of fibers or a gel. \textbf{Figure~\ref{fig: Figure_5}a} shows a spermatozoan swimming through the elongated glycoproteic meshwork in cervical mucus, showing that the obstacle size is much closer to the swimmer size than is often appreciated. Using the pore size as an obstacle length scale, for microorganisms swimming through mucus the Benes number $\mbox{Bn}$ can be in the range $0.1-2$ \cite{pc18}. Similar to the example of shear-thinning by a moving flagellum suggested by \textbf{Figure~\ref{fig: Figure_4}i}, continuum theories using two-phase fluid models suggest that the nature of the network (and so the boundary conditions) on the swimming body are of critical importance for both swimming speed and efficiency \cite{fsp10,dkgf12}. Enhanced motion has been found for undulatory swimmers moving through stiff networks and for magnetically-driven helical bodies swimmers through gels whose meshsize is roughly the same size as the swimmer \cite{smgmmlf14}
(\textbf{Figure~\ref{fig: Figure_5}b,c}). Some organisms even alter the physics of their environment through chemical processes which improve mobility. Urease enzyme released by the pathogen {\it H. pylori} modulates the fluid's local pH but also reduces mucin viscoelasticity, allowing the bacteria to penetrate a mucus lining protecting the gut \cite{ctakgkemse09} (\textbf{Figure~\ref{fig: Figure_5}d}).

\subsection*{Anisotropic fluids} 

\begin{figure}[htbp]
\begin{center}
\includegraphics[width=\textwidth]{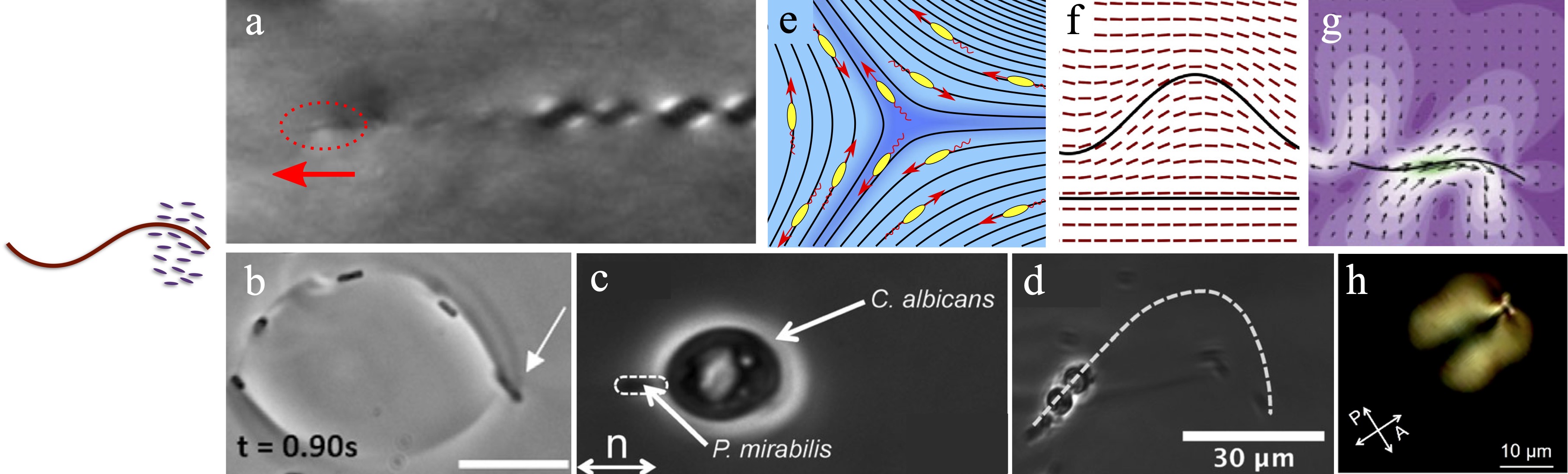}
\caption{Type II systems: anisotropic fluids. (a) Nematic-biphasic melting in the wake of a swimming {\it B. subtilis} cell \cite{zsla14}. {\it Proteus mirabilis} in director-guided motion (b) near a nematic tactoid \cite{mtwa14} and pushing (c) a large yeast cell and (d) two microbeads along prescribed paths \cite{rmasw15}. (e) Depletion of bacteria near a $-1/2$ topological defect \cite{gsla17}. (f) Director field near a waving sheet and an infinite wall in a liquid crystal with strong tangential anchoring conditions \cite{ksp19}, and (g) around a finite body, with velocity field (and magnitude, colored) \cite{lcg21}. (h) A self-propelling liquid-crystal droplet swims through a bulk nematic phase \cite{nca19}. All figures reproduced with permission.}
\label{fig: Figure_6}
\end{center}
\end{figure}

Our last example of a Type II system is swimming in an anisotropic medium which relaxes towards orientational order. Network-formation in mucus can lead to such liquid-crystalline order \cite{vhv93}, and may be relevant to spermatozoa navigation through the cervical canal (\textbf{Figure~\ref{fig: Figure_5}a}) \cite{Chretien03}. A model fluid used to investigate such environments, both experimentally and theoretically, is a liquid crystal. The standard models and fluids used, like the bacteria-friendly lyotropic chromonic disodium cromoglycate (DSCG), are composed of symmetric rodlike molecules, resulting in ordered nematic phases upon cooling, and randomly oriented isotropic phases above a critical temperature. Near the phase transition the motion of a {\it B. subtilis} cell through the environment can temporarily melt the liquid crystal, leaving an alluring helical trail in its flagellar wake (\textbf{Figure~\ref{fig: Figure_6}a}) \cite{zsla14}. 

The details of swimming in a liquid crystal can depend strongly on another feature not present in many other complex fluids, a surface anchoring energy, which penalizes departures from a given molecular orientation there due to surface chemistry. The continuum director field describing the orientation of (stacked) DSCG molecules, for example, preferentially anchors to biological membranes in the tangent plane \cite{mtwa14}. As a consequence, when elongated bacteria are placed in such fluids they tend to align with the local director field - the elastic energy in the bulk liquid crystal is smaller when the body is more ``streamlined''. Exploiting this observation, director-guided motion is seen along the surface of nematic tactoids (\textbf{Figure~\ref{fig: Figure_6}b}) \cite{mtwa14} and has been used to trace a long path for cargo-carrying {\it Proteus mirabilis} cells to follow \cite{rmasw15} (\textbf{Figure~\ref{fig: Figure_6}c,d}). Liquid crystal director fields are also bound by topological constraints, which generically result in the presence of defects, points at which the director field is non-differentiable \cite{dp93}. Different defects induce different bacterial behavior; $+1/2$ defects encourage accumulation while $-1/2$ defects (\textbf{Figure~\ref{fig: Figure_6}e}) result in mean depletion \cite{ptgwl16,gsla17}. 

Theoretical perspectives on swimming in anisotropic fluids are once again stymied by a dramatic coupling of nonlinearities and an avalanche of dimensionless groups, but asymptotic analysis in some simplified settings has provided some insight. The predictions which have emerged are surprising, for instance an infinite swimming sheet can swim either with or against the direction of its wave propagation. The direction depends on the rotational viscosity, which determines the rate at which the fluid's orientational order is recovered, and thus the rate at which an elastically generated ``backflow'' is produced \cite{ksp15,ksp19}. As in the case of isotropic viscoelastic fluids, the bulk elasticity can play the role of effective confinement leading to increased swimming speeds. The degree to which the director field may be deformed is characterized by the Ericksen number $\mbox{Er}=\mu U L/K$, a ratio of the viscous to elastic forces in a flow, where $\mu$ is the fluid viscosity, $U$ and $L$ are characteristic velocity and length scales, and $K$ is a bulk elastic modulus (a Frank elastic constant) \cite{dp93}. Here again, choosing $T_{obs}=\mu L^2/K$ as an elastic relaxation timescale, and $T_{swim}=L/U$ a swimming timescale, the Ericksen number $\mbox{Er}=T_{obs}/T_{swim}$ is the Weissenberg number in different clothing (or the Deborah number in the oscillatory setting). 

The constituents of a liquid crystal generally have a preferred orientation on a given surface, and a surface anchoring energy represents the cost of deviation from this configuration there - large anchoring strengths can distort the field well into the bulk \cite{dp93}. \textbf{Figure~\ref{fig: Figure_6}f} shows the director field of a nematic liquid crystal around an infinite sheet swimming near a flat surface at small Ericksen number, with strong tangential anchoring on both boundaries. While the rotational viscosity and backflow can alter the swimming direction, its effects are substantially magnified when the anchoring strength is large \cite{ksp19}. Simulations of finite-length swimming sheets show similar results but also reveal that the director field deformations and associated elastic stresses, like for swimmers in viscoelastic fluids, are largest near the tail (\textbf{Figure~\ref{fig: Figure_6}g}) \cite{lcg21}. 

Our last example blurs the lines between Type II and Type IV systems. A liquid crystal droplet can swim through a bulk nematic phase due to initial symmetry breaking by Marangoni stresses on the surface, with coherent motion supported by orientation-dependent van der Waals forces at the boundary (\textbf{Figure~\ref{fig: Figure_6}h}) \cite{nca19}. As is generally the case for swimming in anisotropic environments, the swimming direction is biased into the direction of the background director field.

\section*{$\mathcal{III}$. Small $\De$, large $\Bn$, small $\phi$: a single swimmer near a large rigid obstacle}

\noindent {\it There was a soggy collision of a bacterial rod with the ship. The substance of the bacterium bent about the curve of the window, sprang back into shape and bounced off, leaving a smear that washed off slowly.} \cite{akk66}
\vspace{.2cm}

Type III represents systems with a single swimmer in a Newtonian fluid near a rigid obstacle. Depending on the Benes number the obstacle geometry may come into view. Many features of swimming near obstacles, however, is informed by the limit of infinite Benes number (e.g. swimming near an infinite wall). 

\subsection*{Swimming near a wall} 

For very large Benes number the obstacle may appear to the swimmer as a flat wall. One early theoretical study by Katz placed Taylor's infinite swimming sheet model with fixed gait near a rigid boundary \cite{Katz74}. The swimming speed was found to increase dramatically as the body neared the wall but with greater energetic cost. This fact has been used to rationalize swimming enhancement when complex fluid stresses produce effective confinement, as we have already seen in some Type II systems. 

\begin{figure}[htbp]
\begin{center}
\includegraphics[width=\textwidth]{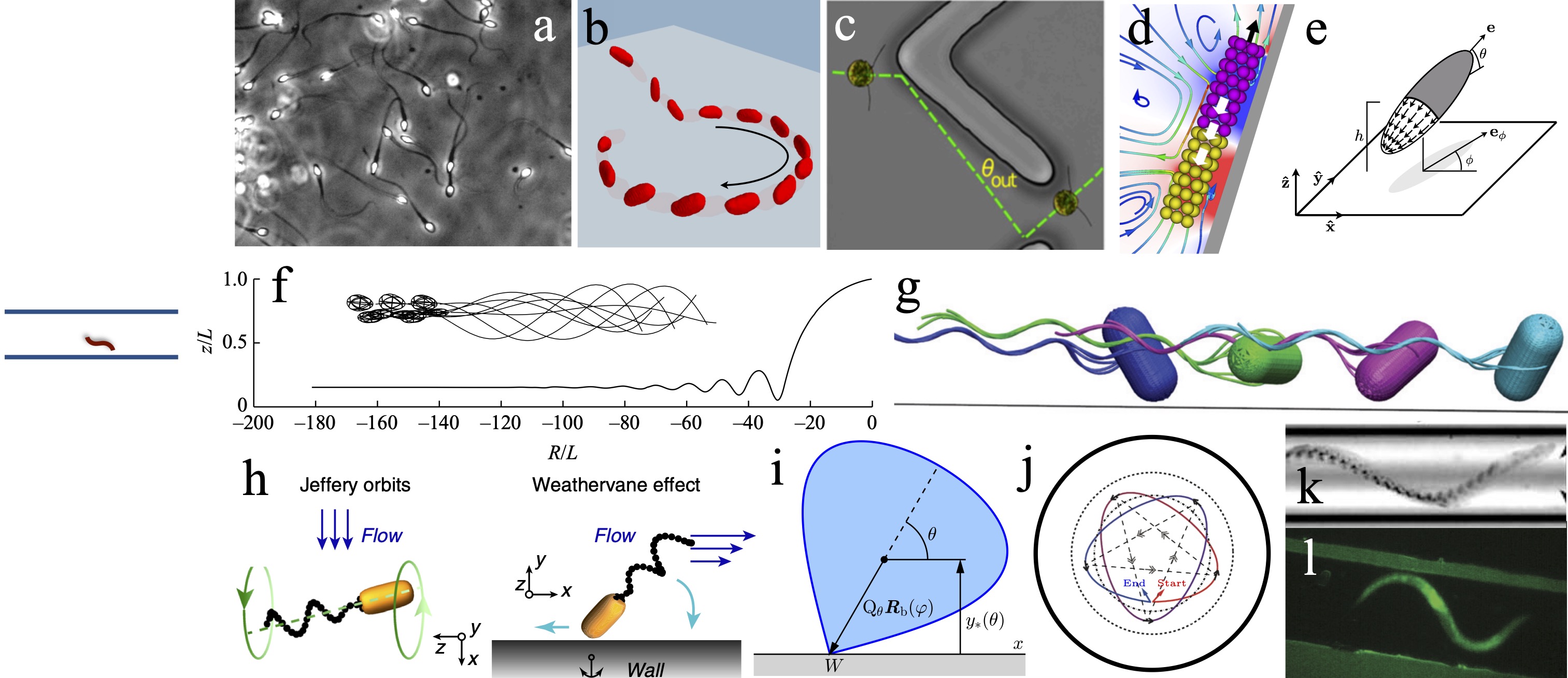}
\caption{Type III systems: swimming near a wall. (a) Surface accumulation of spermatozoa \cite{sb09}. (b) Holographic recovery of circular swimming of an {\it E. coli} cell \cite{bsd17}. (c) A {\it Chlamydomonas} cell departs from a wall at an angle related to its flagellar geometry \cite{kdpg13}. (d) Swimming up a wall against gravity \cite{bulwrwsz21}. (e) Far-field models are often adequately predictive \cite{sl12}, but more detailed simulations can resolve the role of cell shape and deforming bodies, for (f) dynamic accumulation of spermatozoan swimming \cite{sgbk09} and (g) wobbling of multi-flagellated cells \cite{mgr20}. (h) A ``weathervane effect'' due to a background flow can result in upstream swimming \cite{mfjclz19}. (i) Body shape can strongly affect near-wall scattering and statistics \cite{ct21}. (j) Confinement-induced periodic trajectories of a model swimmer \cite{zlb13}. (k) A confined {\it Paramecium} cell swims with a helical trajectory \cite{jus12}. (l) {\it C. elegans} combines swimming and crawling motions under confinement \cite{cfs13}.  All figures reproduced with permission.}
\label{fig: Figure_7}
\end{center}
\end{figure}

More complex body dynamics and trajectories also appear when a boundary is nearby, including surface accumulation \cite{Rothschild63} and circular orbits (\textbf{Figure~\ref{fig: Figure_7}a,b}) \cite{bt90,ffbc95,ldlws06,bsd17}. {\it Chlamydomonas} algae cells scatter from a wall at an angle determined by their flagellar geometry (\textbf{Figure~\ref{fig: Figure_7}c}) \cite{kdpg13,lkg17}, and the random ``tumbling'' of {\it E. coli} is suppressed near surfaces due to increased hydrodynamic resistance \cite{trb00,mbss14}. Hydrodynamic effects can be substantial, metallic synthetic swimmers can even swim upwards against gravity using the hydrodynamic interaction with a wall (\textbf{Figure~\ref{fig: Figure_7}d}) \cite{bulwrwsz21}. Surface accumulation may also be a statement of mean behavior, with individual swimmers undergoing periodic departure and return \cite{bmmt21,ct21}. 

Modeling and simulation has been used to better understand the extent to which hydrodynamics alone can describe these phenomena. Far-field hydrodynamics capture many features robustly. The model Janus swimmer in \textbf{Figure~\ref{fig: Figure_7}e} was used to explore the accuracy of such theories, showing them to be remarkably accurate down even to swimmer-wall distances well below the swimmer length \cite{lp10,sl12}. But to understand the interplay between flagellar beating and surface interactions requires more detailed flagellum-resolving simulations like those shown in \textbf{Figure~\ref{fig: Figure_7}f,g} \cite{gnbnm05,sgbk09,giy10,sgs10,ekg10,mgr20}. The combination of wall effects and a background flow can encourage an unexpected upstream swimming due to a ``weathervane'' effect (\textbf{Figure~\ref{fig: Figure_7}h}) \cite{hkmk07,fmrccal15,ic17,mfjclz19}. Body geometry and fluctuations can also strongly affect orientational statistics \cite{lt09,sgs10,szs15,tpnskrmtgm21}, as elegantly explored in configuration space by Chen \& Thiffeault with the asymmetric model swimmers shown in \textbf{Figure~\ref{fig: Figure_7}i} \cite{ct21}. Other alluring trajectories abound including periodic dynamics under confinement of model swimmers (\textbf{Figure~\ref{fig: Figure_7}j}) \cite{co10,ozm11,zs12,zlb13,swlt17} and in experiments with a {\it Paramecium} cell (\textbf{Figure~\ref{fig: Figure_7}k}) \cite{jus12}. 

Helical bodies in cylindrical confinement enjoy a similar speed enhancement as that described by Katz, though the details depend on the helical geometry and distance to the boundary \cite{lbp14,ls15}. The optimal pitch angle for self-propulsion is similarly affected. The compliance of the cell and particularly any thin, highly deformable motility organelles can thus strongly affect both swimming speeds and energetic costs \cite{lcf19}. Larger slender bodies like {\it C. elegans} also deform under confinement (\textbf{Figure~\ref{fig: Figure_7}l}), and their motility is affected accordingly \cite{lspclhd12,cfs13}.

\begin{figure}[htbp]
\begin{center}
\includegraphics[width=\textwidth]{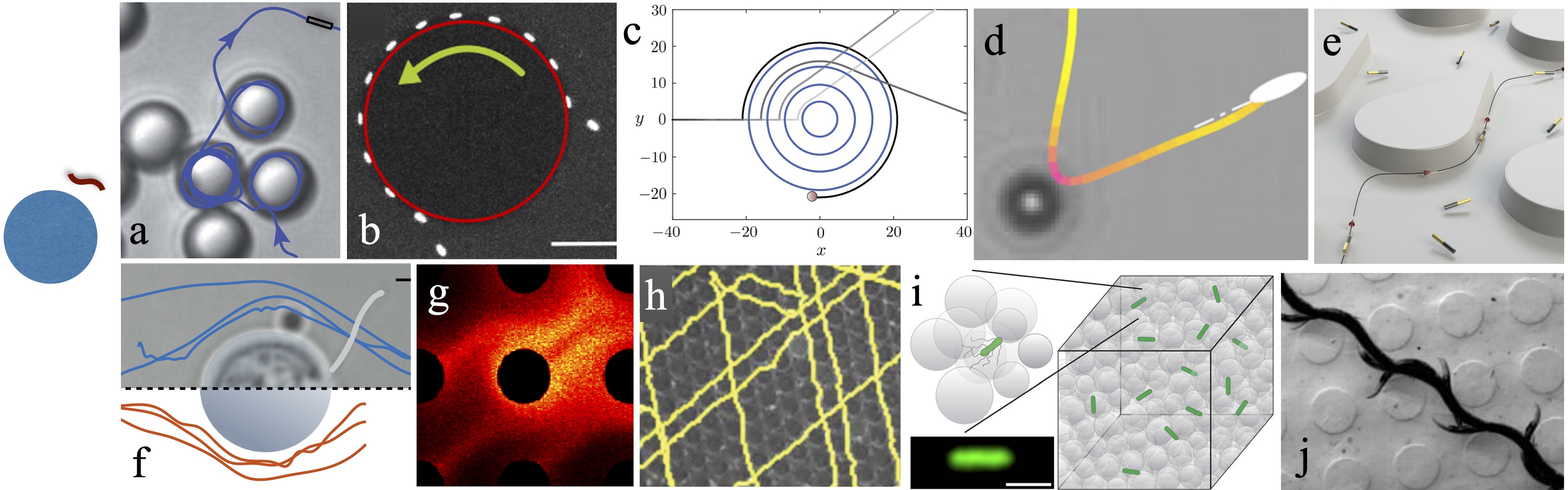}
\caption{Type III systems: geometric obstacles. (a) Scattering and entrapment of synthetic swimmers by spherical colloids \cite{tpbsz14}; and (b) of {\it E. coli} cells by cylindrical posts \cite{sndg15}. (c) Trajectories of pusher-type swimmers around colloids of four different sizes. Scattering is non-monotonic in the colloid size, while entrapment is predicted above a critical Benes number (e.g. colloid size) which depends on the dipole strength and swimmer diffusivity \cite{smbl15}. (d) Cell transport may be enhanced due to scattering events even at low obstacle density \cite{mbacv19}. (e) Scattering may be guided by obstacle shape \cite{wztalrwsz17}. (f) Entrainment of small obstacles depends on their size and the mechanism of propulsion \cite{mjp18}. (g) A simulated cloud of non-interacting swimmers spreading through a periodic array of posts \cite{acs19}. (h) Repeated surface interactions in porous media can focus swimming trajectories \cite{bvdvslp15}; or (i) result in effective diffusion \cite{bd19}. (j) {\it C. elegans} transitions from swimming to crawling in a structured environment to enhance mobility \cite{phnmar08}. All figures reproduced with permission.}
\label{fig: Figure_8}
\end{center}
\end{figure}

\subsection*{Geometric obstacles} 

As the Benes number decreases the obstacle shape begins to come into view ($\Bn=O(1)$ when the body and obstacle are roughly the same size). Billiard-like motion, intermittent periods of entrapped orbiting states, and randomized escape behavior are observed in a fluid with finite-sized obstacles in both synthetic and living systems (\textbf{Figure~\ref{fig: Figure_8}a,b}) \cite{tpbsz14,sndg15}. The far-field flow generated by a non-Brownian pusher-type swimmer (a dipolar swimmer that expels fluid along its swimming direction and draws it in laterally) near a spherical obstacle suggests hydrodynamic entrapment above a critical Benes number which depends on the dipole strength. \textbf{Figure~\ref{fig: Figure_8}c} shows the trajectories of pushers swimming towards colloids of radius 5, 10, 15, and 20 times the body size, showing a non-monotonic scattering angle with increasing colloid size and hydrodynamic entrapment by the largest obstacle \cite{smbl15}. With the inclusion of thermal fluctuations, trapping time statistics depend most notably on the rotational diffusivity of the swimmer. A puller-type swimmer (which instead draws fluid inward along its swimming direction and expels it laterally) can pull itself into an entrapped state near much smaller obstacles. 

If entrapment does not occur or is only temporary, the interaction can still affect the long-time statistics. Such scattering events can enhance transport even at low obstacle densities (\textbf{Figure~\ref{fig: Figure_8}d}) \cite{mbacv19}. The mechanism of self-propulsion \cite{smbl15,weg15,cltkp15,mskm16,lbsm16} and obstacle shape \cite{gkca07,wztalrwsz17}, for instance the teardrop-shaped obstacles shown in \textbf{Figure~\ref{fig: Figure_8}e}, can also strongly affect the interaction with and departure from the surface. Sorting and rectification devices meant to exploit such interactions have been constructed and explored, including funnels and gears \cite{bdlrvv16}.



For smaller obstacles it becomes more important whether or not they are fixed in space or free to become entrained by a passing swimmer. If entrained, the shape and stroke of the swimmer become particularly important (\textbf{Figure~\ref{fig: Figure_8}f}) \cite{mt17,mjp18}. Transport of yet smaller obstacles, or even fluid particles, return us closer to Type I problems and the general question of fluid entrainment by swimming bodies \cite{ltc11}. A direct exploration of the Benes-number dependence on entrapment and particle entrainment is found in Ref.~\cite{sy17}, and a generic enhancement of swimming by the reduction of wobbling in colloidal suspensions is described in Ref.~\cite{kslfxc22}.

There may instead be many large obstacles to navigate, and the dynamics described above may take place over repeated interactions. \textbf{Figure~\ref{fig: Figure_8}g} shows a simulated cloud of active Brownian swimmers spreading in a periodic array of posts. Generic features in such systems include boundary layers and thin regions of increased swimmer concentration in the ``wake'' of an obstacle \cite{acs19}. In systems with more tightly packed obstacles, synthetic swimming particles can hop from colloid to colloid with a trapping time that depends on fuel concentration, while {\it E.~coli} trajectories are rectified into long, straight runs (\textbf{Figure~\ref{fig: Figure_8}h}) \cite{bvdvslp15}. The uncertainty in trapping times and exit orientations leaving one obstacle leads to an effective diffusion through a bulk porous medium (\textbf{Figure~\ref{fig: Figure_8}i}) \cite{bd19}. In this example, depending on whether the obstacle size is chosen to be the size of the colloid, roughly 10$\mu$m, or the narrow pore size between obstacles, $10^{-1}\mu$m, the Benes number ranges from $10^{-2}$ to $1$.

As usual body deformability can be important; {\it C. elegans} for instance transitions from swimming to crawling motions to enhance its mobility in a structured environment. \textbf{Figure~\ref{fig: Figure_8}j} shows this nematode traversing a periodic array of obstacles. When $\mbox{Bn}$ is order 1, particularly if the undulatory wavelength is the same size as the obstacles, the swimming is focused into straight trajectories like the {\it E.~coli} cells in \textbf{Figure~\ref{fig: Figure_8}h} as the body pushes off each obstacle one after another \cite{phnmar08,mkzs12}.

\section*{$\mathcal{IV}$. Large $\De$, large $\Bn$, small $\phi$: a single swimmer near a deformable obstacle}

\noindent {\it The white cell was tremendous. It was five times as large in diameter as the Proteus, perhaps larger; a mountain of milky, skinless, pulsing protoplasm in comparison to the individuals watching.} \cite{akk66}
\vspace{.2cm}

\begin{figure}[htbp]
\begin{center}
\includegraphics[width=\textwidth]{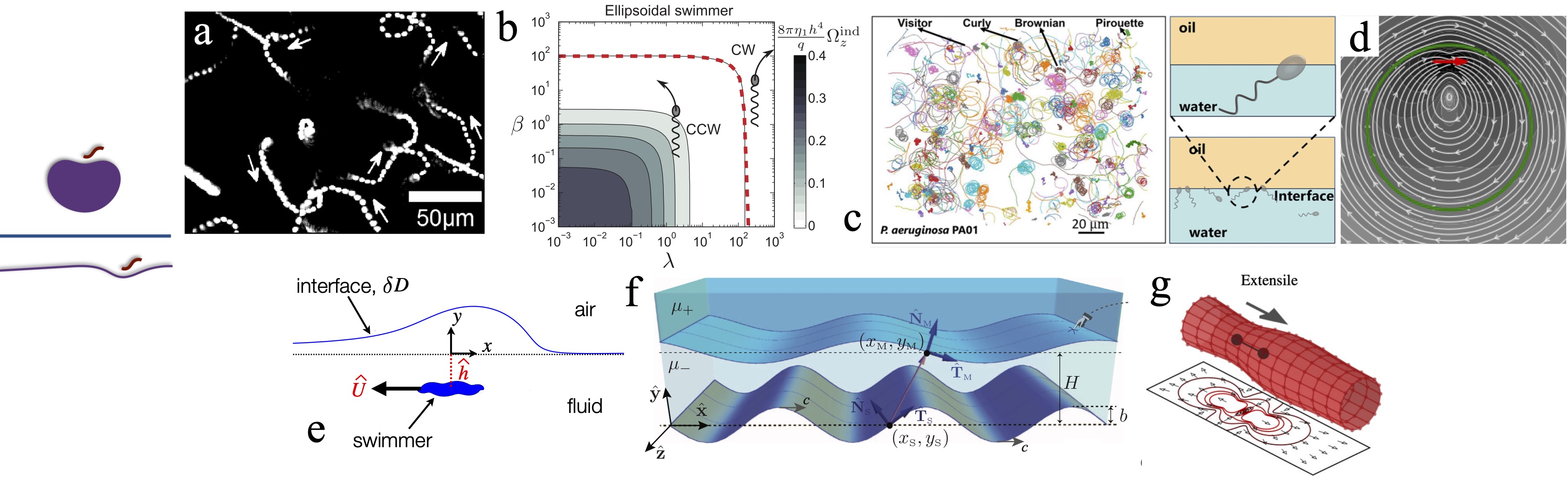}
\caption{Type IV systems. (a) Counter-clockwise circular trajectories of {\it C. crescentus} cells at an air-water interface~\cite{mhlmt13}. (b) Purely hydrodynamic arguments show similar counter-clockwise (CCW, free surface) to clockwise (CW, solid wall) rotations depending on the viscosity ratio $\lambda$ and the surface viscosity $\beta$ (contours of the rotation rate are shown) \cite{ll14}. (c) Free and adhered states of {\it {P}. aeruginosa} cells at an oil-water interface - distinct trajectories (left) are due to differing trapped states (right) \cite{dmcs20}. (d) Flows in and along the surface of a confining droplet due to an interior point force \cite{ssalmglmd20}. (e) Normal surface deformations may be generated by distant swimmers \cite{clslh11}. (f) Closer swimmers can deform an interface more substantially, resulting in stronger speed and efficiency adjustments \cite{dp13}. (g) Soft confinement can either increase or decrease the speed of model swimmers \cite{ly13}. All figures reproduced with permission.}
\label{fig: Figure_9}
\end{center}
\end{figure}


In Type IV systems a large obstruction relaxes on a timescale longer than or commensurate with that of a single organism's motion. Examples include air-water interfaces, oil-water interfaces \cite{Conrad20}, thin films \cite{mdys16}, and mucus-lined tissues like those found in the mammalian reproductive tract \cite{sp06}. Even if the action of a swimmer is insufficient to drive substantial normal surface deformations, the freedom of tangential interface motion can affect swimming behaviors. For example, the orientation of the circular swimming patterns of organisms like {\it E. coli} are reversed relative to what is observed for swimming near a rigid wall (\textbf{Figure~\ref{fig: Figure_9}a}) \cite{lpcp10,mhlmt13,ll14,bsfdd19}. This change in observed behavior can be rationalized by examining the hydrodynamics of swimming with a mirror image across the interface (\textbf{Figure~\ref{fig: Figure_9}b}) \cite{lpcp10,dldaai11,sl12,ll14,pcgc16}. But the behavior can also be very sensitive to the surface adsorption of organic materials and surfactants, which can render the surface more or less rigid \cite{mhlmt13}. 

Surface absorption of the swimmer is possible as well, which can lead to distinct trajectories depending on the nature of the trapped state. \textbf{Figure~\ref{fig: Figure_9}c} shows numerous trajectories of {\it {P}. aeruginosa} cells at an oil-water interface, which include diffusive paths, curved trajectories, and rapid ``pirouette'' rotations \cite{dmcs20}. The tangential flows and resultant swimmer dynamics generally depend on the relative obstacle size or interfacial confinement and the common possibility that surfactants are present on the interface (\textbf{Figure~\ref{fig: Figure_9}d}) \cite{dsa18,ssalmglmd20}. Surfactants can rigidify the interface and nudge the system back towards Type III interactions of a swimmer with a rigid obstacle or wall \cite{sa19,awva19}. 

Stronger swimmers, or softer boundaries, can result in normal surface deformations as well. Many theoretical techniques have been used to probe this problem including far-field approximations (\textbf{Figure~\ref{fig: Figure_9}e}) \cite{tyhl08}, a conformal mapping approach \cite{clslh11}, and small amplitude asymptotic analysis for swimming sheets (\textbf{Figure~\ref{fig: Figure_9}f}) \cite{dp13} and squirmers \cite{sa17}. Depending on the nature of the swimmer and the surface physics, the deformability of the boundary can result in either enhancement or retardation of swimming speeds for a given swimming kinematics (\textbf{Figure~\ref{fig: Figure_9}g}) \cite{dp13,ly13}. Even tighter confinement can lead to even more dramatic boundary and swimmer shape deformations. A model amoeboid swimming through a narrow compliant channel, for instance, undergoes a marked decrease in swimming speeds as the channel narrows \cite{wthfrlm15,dfm20}.

\section*{$\mathcal{V}$. Small $\De$, small $\Bn$, large $\phi$: many swimmers in a suspension of small rigid obstacles}

\begin{figure}[htbp]
\begin{center}
\includegraphics[width=\textwidth]{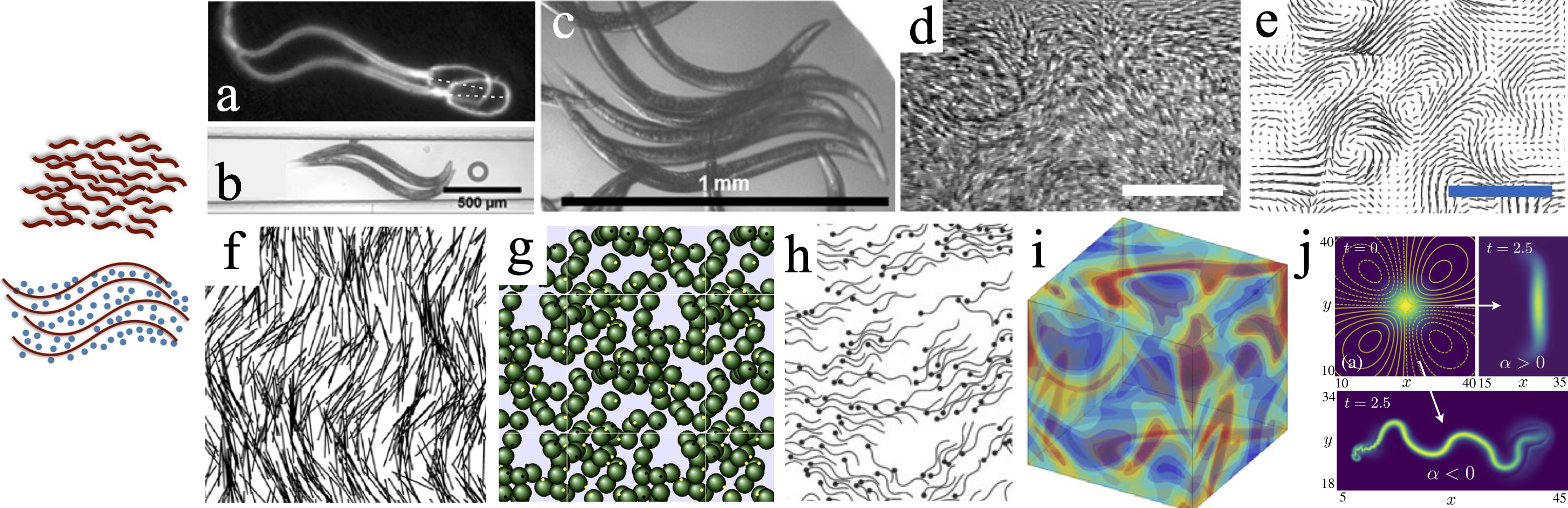}
\caption{Type V systems. (a) Synchronization of swimming spermatozoa \cite{wcgr09}; (b,c) and {\it C. elegans} \cite{yrb14}. (d) Collective motion of {\it B. subtilis} cells shows correlations on much larger scales than the individual swimmer size (scale bar is 35$\mu$m) \cite{dccgk04}. (e) Velocity field from collective simulations resolving individual swimmers (scale bar is 5 swimmer lengths) \cite{hsg05}. (f) Onset of an orientational instability in a suspension of nearly aligned ``pusher'' swimmers \cite{ss07}. (g) Collective motion of model spherical squirmers \cite{eiyl11}. (h) Synchronization and aggregation within large groups of model spermatozoa \cite{sk18}. (i) Swimmer concentration field via solution of moment equations in a coarse-grained theory \cite{ss13b}. (j) Self-stretching of a Gaussian concentration of rightward-pointing swimmers, along the swimming direction and unstable for pushers $(\alpha<0)$ or transverse to the swimming direction for pullers  $(\alpha>0)$ \cite{mess19}. All figures reproduced with permission.}
\label{fig: Figure_10}
\end{center}
\end{figure}

Type V systems feature environments with small rigid obstacles, as in Type I, but with a sufficiently large swimmer concentration $\phi$ so that their interactions become relevant. With $\Bn$ small the fluid is again reasonably approximated as a continuum medium at the scale of the swimmers. A first step in the direction of understanding large swimmer concentrations is to consider the interactions of two swimmers. When two swimmers approach each other the details of their shapes and swimming cycles come into view. For example, waving sheets and filaments can be synchronized through hydrodynamic interactions - \textbf{Figure \ref{fig: Figure_10}a} shows two nearby swimming spermatozoa swimming in synchrony \cite{wcgr09}. But hydrodynamic interactions are not always dominant in such settings. The synchronization of {\it C. elegans} nematodes, for instance, is based more on the effects of steric hindrance (\textbf{Figure \ref{fig: Figure_10}b,c}) \cite{yrb14}.

The collective dynamics of many swimmers, meanwhile, remains a particularly vibrant research area which has been treated to excellent and deeper treatments beyond the scope of this review \cite{Ramaswamy10,ks11,vz12,mjrlprs13,ewg15,ss15,acj22,bfmr22,ssbmv22}. Among the main features observed in these systems, the long-range nature of hydrodynamic interactions in viscous fluids can result in positional and orientational correlations among groups of swimmers even at volume fractions as low as $\phi=1-2\%$  \cite{dccgk04,grlc14}. Such correlations can depend on body shape - rotation and alignment is different for sphere-like and rod-like bodies \cite{kk13}. At large swimmer concentrations, new complex phenomena begin to emerge. Swimmer interactions can lead to increases in effective velocity and diffusion. For example, \textbf{Figure \ref{fig: Figure_10}d} shows a suspension of {\it B. subtilis} cells, each roughly 4$\mu$m long, with structures observed on the larger scale of the 35$\mu$m bar on the bottom right \cite{dccgk04}. These structures lead to enhanced diffusion of tracer particles that depend on their size \cite{pgpa16} and shape \cite{pltzxc16}. \ch{Interactions among externally driven rotating bodies can result in clustering, colony-scale rotations, and so-called odd viscosity \cite{sbmsbsi19,hktgas20,ssbmv22}. Such dynamics have also been observed in swimming suspensions of {\it T. majus} bacteria cells \cite{pwl15} and starfish embryos \cite{tmlchfgdf22}.}


\subsection*{Simulations and theoretical advances}

A number of agent-based simulations have been used to explore the mechanisms causing the group behavior. Hernandez-Ortiz et al. \cite{hsg05} showed that the type of swimming (either pushed from the back or pulled from the front) changes the hydrodynamic interactions and thereby the large scale group behavior. As in experiments, the velocity field shown in \textbf{Figure \ref{fig: Figure_10}e} shows correlated motion on a scale much larger than the individual swimmer size; the included scale bar is 5 swimmers long. A generic feature in active suspensions for small $\phi$ is the emergence of long-range orientational order for pusher-type swimmers, seen for instance in simulations of many pusher-type swimming rods (\textbf{Figure \ref{fig: Figure_10}f}) \cite{ss07}. Such group behavior is found in simulations using spherical squirmers as well (\textbf{Figure \ref{fig: Figure_10}g}) \cite{eiyl11}. In this system at higher swimmer concentrations the isotropic state destabilizes and relaxes into polar order for both pushers and pullers, unlike in dilute theories for which only isotropic suspensions of pushers are orientationally unstable.

These agent-based simulation models have quantified how hydrodynamic interactions and swimmer shape impact swimmer velocity and orientational correlations \cite{ss07,uhg08,ilp08,ug11,ss12,bnnms19}. They have also used passive tracer particles which are advected and mixed by the flows created by the swimmers as a method to quantify collective behavior \cite{kkw14,mm14,snnmm17}. Within large groups, clustering is observed, and is attributed to steric and hydrodynamic pairwise interactions similar to those seen in dilute suspensions (\textbf{Figure \ref{fig: Figure_10}h}) \cite{sk18}. Such results highlight one feature that is currently less understood: how the details of swimmer shapes and their swimming strokes affects short range correlations within systems that show large scale group dynamics \cite{bbs15,shk20}.

Theoretical advances have involved the derivation and analysis of continuum field equations for the coarse-grained system dynamics \cite{ss15}. Further simplification is found through closure approximations, generating coupled equations for the first few local orientational moments, which produce results similar to those found in full agent-based simulations. \textbf{Figure \ref{fig: Figure_10}i} shows spatial variations in a concentration of pushers which have emerged out of an isotropic state using the Saintillan-Shelley model equations \cite{ss13b}. The self-stretching of localized colonies, and a cascade of transverse instabilities in thin concentration bands, provides insight on these more fully developed global dynamics (\textbf{Figure \ref{fig: Figure_10}j}) \cite{sr02,mess19}. Recently, field theories that include fluctuations and correlations have been developed to examine the onset of collective behavior \cite{qku17,snnmm17,snsmm20}.

\subsection*{Complex rheology}

The activity and orientational order which emerges in these systems can confer the solvent-suspension system with features normally associated with bulk complex fluids. The effective viscosity of fluid containing puller-type swimmers, specifically {\it Chlamydomonas Reinhardtii}, has been found to increase on account of their swimming activity \cite{rjp10,mrpw13}. Suspensions of pusher-type swimmers like {\it B. subtilis} \cite{sa09} and {\it E. coli} \cite{gmblrc13,lgdc15,mgloyfp15}, meanwhile, are found to decrease the effective viscosity of the medium. These generic effects of extensile (pusher) and contractile (puller) suspensions are also recovered theoretically \cite{hrrs04,Saintillan10}. As the swimmer volume fraction increases the decreased viscosity for suspensions of pusher-type swimmers can reach zero, effectively a superfluid state, where the medium can be sheared with no stress, and then into negative effective viscosity, where work must in fact be done to keep the medium from shearing itself \cite{hrrs04, ip07, habk08, sa09,Saintillan18}. The superfluid response can also cause shear-banding in the velocity profile \cite{gspxc18}.

These suspensions are often found to have a swimmer contribution that decreases with background shear rate, as the active stresses which they produce become drowned out by the large scale motion. Many models have computed the contribution to rheological properties from the hydrodynamic disturbance caused by the swimmers. These models show a decrease due to shear-induced swimmer alignment, much like the shear-thinning found in suspensions of passive rods. There is also a diffusive contribution, which can lead to non-Newtonian behavior of the suspension \cite{tb17}. 

Suspensions of spherical squirmers have been used to explore the rheology of very high swimmer concentrations, where the lubricating region between individual swimmers dominates their interactions. The effective shear viscosity is found to increase for both pushers and pullers compared to a similar suspension of inert spheres. But for bottom-heavy squirmers the distinction between pushers and pullers returns, and viscosity reductions down to superfluidity appear again for pusher suspensions \cite{ibp21}. 

The role of body deformability on active suspensions and their effective rheology remains a generically open area of inquiry, but simulations have revealed that the sign of the effective medium's second normal stress differences depends on body stiffness \cite{moi20}. Other non-Newtonian features due to active suspensions like viscoelasticity \cite{hrrs04,Saintillan10,bu14b,bk17}, and yield-stress behavior \cite{cfmoy08,glm10} have also been predicted. The influence of surfaces described in Type III can also impact measurements of rheological properties \cite{lzc19}. For a more detailed analysis of active suspension rheology in general see the recent review by Saintillan \cite{Saintillan18}. 

Type V systems include another fairly unexplored territory, active suspensions in media with somewhat larger obstacles (though still smaller than the swimmer size). If those small stiff obstacles are free to move, they may simply alter the viscosity of the suspending fluid. But if the obstacles are fixed either through connections to each other or to the surroundings, they will alter the swimmer behavior in a different way. The stiff connection allows for the dissipation of momentum and the obstacles will generically screen long-ranged hydrodynamic interactions. Instead, short ranged interactions can lead to large scale patterns \cite{ha10}.

\section*{$\mathcal{VI}$. Large $\De$, small $\Bn$, large $\phi$: many swimmers in a suspension of small deformable obstacles}

\noindent {\it ``If you damaged it, chemicals were released into the blood-stream; chemicals that attracted white cells from all the neighboring regions.'' ``Then, for God's sake, swim!''} \cite{akk66}
\vspace{.2cm}

Type II systems were characterized by small obstacles with intrinsic timescales longer than or commensurate with that of a single swimming body. If there are instead many interacting swimmers in such an environment we find ourselves in a Type VI system. The group of swimmers may be invading a shear-thinning viscoelastic solution or gel, or an anisotropic medium like mucus. 

Depending on the system, the non-Newtonian bulk fluid features may have a bearing primarily on localized behaviors and interactions, or may reveal structures on much longer length scales than the swimmers themselves. The motion of each swimmer can be changed, as previously described in Type II, and in some settings these changes to individual swimming behaviors might be sufficient for understanding the effect of viscoelasticity on group behavior. But the hydrodynamic interactions may also be modified to produce a group behavior that is not simply a shift or rescaling of theories based on Newtonian fluids.

\begin{figure}[htbp]
\begin{center}
\includegraphics[width=\textwidth]{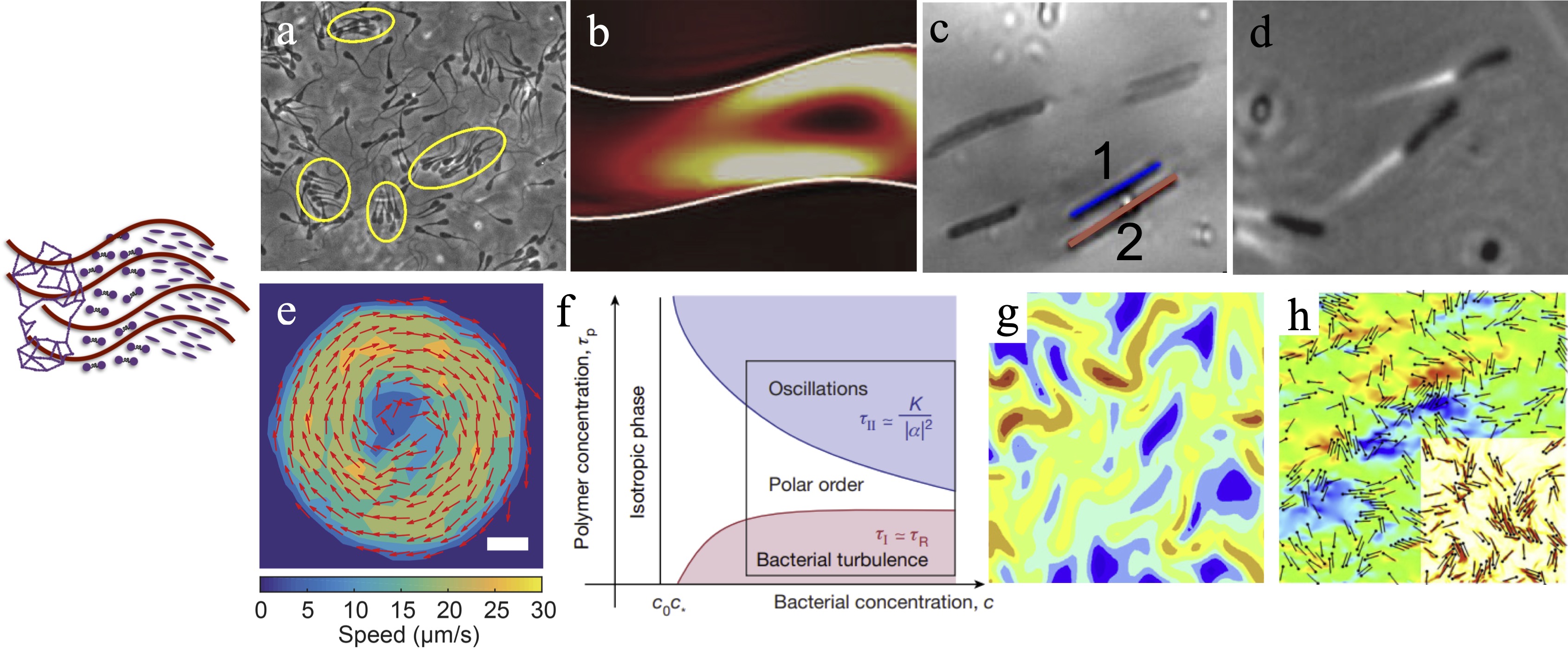}
\caption{Type VI systems. Interacting swimmers in a fluid with bulk complex features. (a) Viscoelasticity-enhanced synchronization in spermatozoa \cite{tlhfaws17}; (b) and in simulations of swimming sheets \cite{cfs13}. \ch{(c) Swimming speeds of nearby {\it B. subtilis} cells converge in an anisotropic medium \cite{szla15}. (d) Surface anchoring induces bacterial collision and group swimming \cite{ztgsal17}.} (e) Velocity field and magnitude of an active suspension of {\it E. coli} cells in a viscoelastic medium; and (f) associated phase diagram of ordered and disordered states, including direction-reversing oscillations above a critical polymer concentration \cite{lsmw21}. (g) Swimmer concentration from mean-field simulations of an active suspension in a viscoelastic medium reveal a reduction of large-scale structures \cite{bu14}. (h) Agent-based simulations show a similar reduction of large-scale structures \cite{la16}. All figures reproduced with permission.}
\label{fig: Figure_11}
\end{center}
\end{figure}

As the volume fraction of swimmers $\phi$ increases, the first deviation from individual swimmer dynamics occurs due to pair interactions. Viscoelastic fluids lead to enhanced clustering and synchronization in experiments with spermatozoa (\textbf{Figure \ref{fig: Figure_11}a}). Infinite swimming sheets synchronize in polymeric fluids much more rapidly than they do in a Newtonian fluid (\textbf{Figure \ref{fig: Figure_11}b}) \cite{epl10,cfs13}. \ch{Liquid crystals also alter the strength and anisotropy of pair interactions, which can result in the convergence of swimming speeds (\textbf{Figure \ref{fig: Figure_11}c}) \cite{szla15}. The interaction may depend strongly on surface anchoring conditions on the swimmers and on nearby boundaries. In some cases the liquid crystal can contribute to swimmer repulsion; in others, collisions and group swimming have been observed (\textbf{Figure \ref{fig: Figure_11}d}) \cite{ztgsal17}.}

For larger swimmer volume fractions, group behavior on length scales much larger than the individual swimmer size once again emerges as in Type V systems. Changes in the velocities within large groups have been observed experimentally. \textbf{Figure \ref{fig: Figure_11}e,f} show the instantaneous velocity field of a suspension of {\it E. coli} in a viscoelastic medium, and a phase diagram of ordered and disordered states across different bacterial and polymer concentrations. For sufficiently large polymer concentration a bizarre effect is observed, the onset of occasional group transport direction reversal \cite{lsmw21}. Inhibition of large scale group behavior and an increase in swirling oscillations are attributed to a interaction between the time scale of the collective vortex flow and the time scale of elastic fluid relaxation. This example sits at the interface between Type VI and, say, Type VII, as the system behavior depends upon both confinement and bulk rheology, e.g. upon multiple obstacles on multiple scales. The question of classification in any system requires identification of the dominant contribution for the features of interest. 

The earliest works to examine these effects theoretically used a continuum field theory to track the dynamics of the active particles and a continuum stress field to track the dynamics of the complex fluid  (\textbf{Figure \ref{fig: Figure_11}g}) \cite{bu11,bu13,bu14}. A prototypical view of group behavior among pushers is as a competition between hydrodynamic interactions that lead to alignment of swimmers and noise that tries to randomize swimming directions. Because the viscous response is driving an instability to group behavior, if the fluid does not fully relax during the instability, a viscoelastic fluid produces less group behavior than a Newtonian fluid with the same zero-shear viscosity. However, because the suspended soft obstacles do contribute to the viscous response, there can be more group behavior than if the obstacles were not present.

In addition to changing fluid instabilities, viscoelastic fluids can change the nature of group behavior. This group behavior is often characterized by chaotic swirling flows. Viscoelastic stresses can break up these large swirls. This leads to intermittancy similar to turbulent drag reduction; a random isotropic suspension can be unstable, leading to large-scale flows which excite viscoelastic stresses - stresses which proceed to suppress the large-scale flows. Other works modeling group behavior in viscoelastic fluids have tracked interacting swimmers within a continuum viscoelastic fluid (\textbf{Figure \ref{fig: Figure_11}h}) \cite{laprl16}. Such models also show a suppression of large scale group behavior and fluctuations in viscoelastic fluids. Alternatively, in liquid crystals with homeotropic anchoring, distortions of the director can mediate swimmer interactions and cause collective effects \cite{ztgsal17}.

Some soft systems do not fully relax at long times due to cross-linking of polymers, as in a gel. These properties have the potential to alter the behavior of concentrated groups. For example, mucus gels can change the swimming and aggregation of bacteria \cite{cfkafkr12}. One important feature of such systems is that the swimmers can change their behavior in response to the mechanical properties as well as chemical makeup of the suspending material \cite{wwr18}.

\section*{$\mathcal{VII}$. Small $\De$, large $\Bn$, large $\phi$: many swimmers near a large rigid obstacle}

\noindent {\it ``They won't bother us, I hope. Any bacteria in the circulatory system reaches a lymph node eventually. It can't negotiate the narrow twisting channels...''. ``Can we?''} \cite{akk66}
\vspace{.2cm}

\begin{figure}[htbp]
\begin{center}
\includegraphics[width=\textwidth]{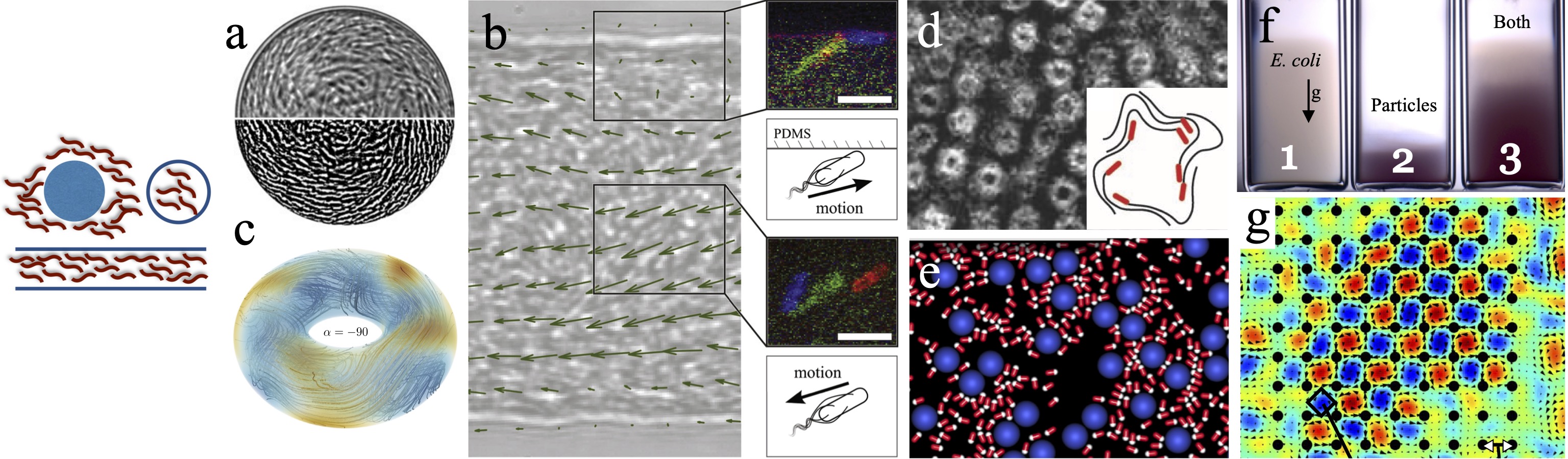}
\caption{Type VII systems. (a) Edge rotation and bulk counter-rotation due to confinement in a circular domain \cite{lwg14}; (b) and in a racetrack, where swimmers in the bulk are carried opposite their orientation by a flow generated by swimmers at the boundary \cite{wlg16}. (c) Confinement-induced organized swimming in three-dimensions in a coarse-grained model \cite{ts19}. (d) Vortices of swimming spermatozoa near a flat surface \cite{rkh05}. (e) Active particles lead to attraction among passive colloids akin to a depletion effect \cite{ambrd11}. (f) Activity by swimming {\it E. coli} reduces the sedimentation speed of passive particles \cite{spmpa21}. (g) A lattice of bacterial vortices organized by an array of periodic obstacles \cite{rnhsbka20}. All figures reproduced with permission.}
\label{fig: Figure_12}
\end{center}
\end{figure}

The collective dynamics of swimming organisms can be strongly influenced by confinement, which brings us to Type VII systems. Swimming bodies at a surface are more restricted in the orientations that they may exhibit, as seen in Type III systems with single swimmers. At the group level such restrictions can propagate into and influence the bulk suspension organization. For example, suspensions of confined bacteria can settle into a large-scale group rotation (\textbf{Figure~\ref{fig: Figure_12}a}) \cite{lwg14}. When multiple rotating domains are connected by channels, the direction of bulk flow which is normally selected with equal probability may start to become influenced by neighboring domains, resulting in the possibility of ferromagnetic or antiferromagnetic global states \cite{wwdg16}. But the relationship between individual body motion and the motion of the group is is not always simple. The flagellar motion of cells swimming near the boundary in \textbf{Figure~\ref{fig: Figure_12}b} generates a flow in the opposite direction in the bulk, so confinement simultaneously produces both upstream and downstream bacterial transport \cite{wlg16}. 

Just as in Type V systems, coarse-grained modeling has identified relationships between the geometry of confinement and the strength of the far-field hydrodynamic signal (dipole strength). The interplay between them can result in coherent rotational motion, or deterioration into other classical aperiodic roiling states, or dynamic modes combining features of both (\textbf{Figure~\ref{fig: Figure_12}c}) \cite{ss15,tas17,ts19}. Directed motion is found when the channel width is close to the length-scale of rotating domains in the unbounded system \cite{wlg16}. With even tighter confinement, phase transitions to jammed states have been found, explored using suspensions of model spherical squirmers \cite{zs14}. 

A boundary may encourage more localized collective behavior as well. Suspensions of spermatozoa near a wall, for example, settle into an array of swimming vortices (\textbf{Figure~\ref{fig: Figure_12}d}) \cite{rkh05}. But while confinement can encourage the emergence or selection of collective behavior, the screening of flows by nearby boundaries can also inhibit the onset of large scale structures. This was shown numerically in Ref.~\cite{houg09} using a suspension of model swimmers composed of dipolar pushing beads. Screening effects may be partially offset by a stronger hydrodynamic signal, external forces or torques on immersed bodies can generate different hydrodynamic signals as viewed far from afar. A suspension of active rollers, for instance, settles into smaller rolling groups due to long-ranged, wall-mediated interactions \cite{ddysdc17}. \ch{If the rotations are instead about the wall normal, confinement can lead to peculiar effects like topologically protected edge currents and colony-scale oscillations \cite{vpibv16,dmv18,yrcz20,lzzdnwclzyf20,zhsss20}.}

New features appear as the obstacles shrink in relative size to the scale of the swimmers themselves. Active particles can produce effective attraction between passive colloids similar to a depletion effect (\textbf{Figure~\ref{fig: Figure_12}e}) \cite{ambrd11}. The sedimentation speed of passive particles can be reduced by swimming bacteria, seen in experiments with swimming {\it E. coli} cells and polystyrene beads (\textbf{Figure~\ref{fig: Figure_12}f}) \cite{spmpa21}. If the obstacles are fixed in space they can encourage the development of bacterial vortices just as was the case with external confinement (\textbf{Figure~\ref{fig: Figure_12}g}). With certain obstacle arrangements (Kagome lattices) such obstacles can even result in net colony rotation \cite{rnhsbka20}. 


\section*{$\mathcal{VIII}$. Large $\De$, large $\Bn$, large $\phi$: many swimmers near a large deformable obstacle}

In Type VIII systems the deformation of large boundaries are again substantial to the point that they can affect the behavior of a suspension of swimming cells. We have already seen that confinement of active suspensions can encourage the emergence of directed motion of a bulk suspension. But when the confinement is deformable, the relaxation timescale and associated surface tractions are strongly coupled to the dynamics of the suspension. Among other physical phenomena this territory includes moving droplets of active matter. 

\begin{figure}[htbp]
\begin{center}
\includegraphics[width=\textwidth]{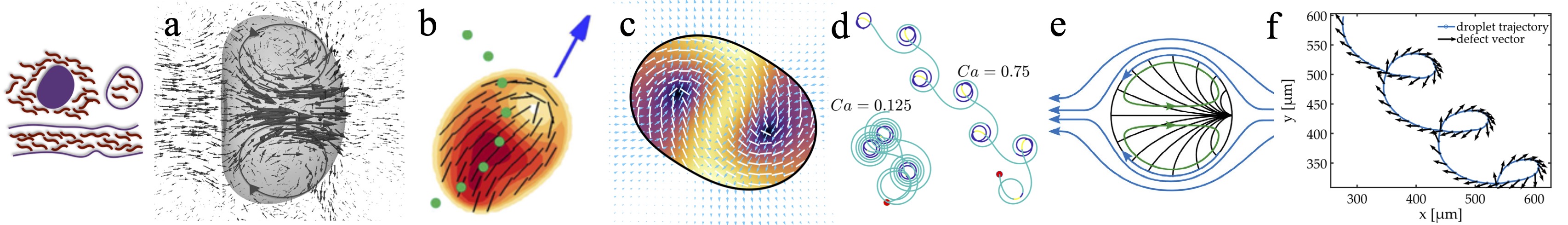}
\caption{Type VIII systems. (a) A broken symmetry with non-motile stress-generating particles under soft confinement leads to swimming \cite{tmc12}. (b) Zigzagging of an active nematic droplet \cite{gl17}. (c,d) Periodic wobbling and a transition to swimming modes of a coarse-grained non-motile suspension, tuned by the capillary number $\mbox{Ca}$ \cite{yss21}. (e,f) Marangoni stresses on the surface of a liquid-crystal droplet couple to internal elastic modes resulting in coherent swimming, while additional symmetry-breaking results in chiral trajectories \cite{kkbm16}. All figures reproduced with permission.}
\label{fig: Figure_13}
\end{center}
\end{figure}

As in Type IV systems some of the most common deformable boundaries are fluid-air and fluid-fluid interfaces. When sufficiently many bacteria adhere to an interface they may form biofilms, introducing to the interface elastic properties which depend on the strain of bacteria \cite{vmnlls17,dsa18}. Bacteria-laden interfaces can also play an emulsion-stabilizing role similar to surfactants \cite{dyfg04} and may render the surface viscoelastic \cite{rhsl21}. Once a biofilm is formed the relevance of swimming diminishes, but the behavior of planktonic (freely swimming) bacteria near these surfaces can impact its spread into new territories and whether the biofilm forms at all \cite{Mazza16}. Such biofilms can disrupt natural medical function, for instance in persistent infections like endocarditis \cite{csg99}, or can be useful for applications such as bioremediation and biodegradation. 

More recent advances have considered the captivating dynamics of active droplets: active suspensions with compliant, mobile confinement. Non-motile suspensions of particles which exert an extensile stress on the surrounding fluid are known to produce large-scale flows and features similar to the swimming systems described in Type V \cite{scdhd12}. But when confined inside a droplet the resulting interaction with the boundary can lead to geometric symmetry breaking and swimming motion (\textbf{Figure~\ref{fig: Figure_13}a}) \cite{tmc12}. More intricate dynamics are accessed by tuning the relative size of the droplet; and active nematic model showed periodic changes in alignment and stretching directions resulting in zigzag trajectories  (\textbf{Figure~\ref{fig: Figure_13}b}) \cite{wh16,gl17}. A coarse-grained model was used to reveal additional contributions of internal director field oscillations, and an array of different trajectories depending on the size of the surface tension relative to the internal stress contributions (\textbf{Figure~\ref{fig: Figure_13}c,d}) \cite{yss21}.

Active droplets present a curious question of placement for the classification system, as will always be the case when multiple important scales are present. When droplets move as a consequence of active internal stresses, what is more central, the self-transporting suspension dynamics at the microscale, or the net effect resulting in what appears to be a single swimmer at a much larger scale? We have included these systems in Type VIII due to the critical importance of the boundary deformability, but in this case the boundary might be considered the swimmer itself. \textbf{Figure~\ref{fig: Figure_13}e} shows a schematic for how Marangoni stresses on the surface of a liquid crystal droplet can couple to internal elastic modes, resulting in coherent swimming, along with observed chiral trajectories (\textbf{Figure~\ref{fig: Figure_13}f}) \cite{kkbm16}. For a broader discussion of active droplets see the review by Maass et al.~\cite{mkhb16}.

\section{SUMMARY AND OUTLOOK}

We have described a proposed classification for swimming in complex fluids. The organization is based on the relationship of a swimming body to one or many obstacles in a fluid, broadly interpreted, be they molecular fluid constituents or enormous confining boundaries. The relative length and time scales of the swimmers and obstacles, along with the concentration of swimmers, impacts the key physical principles at play and behaviors observed. We hope that this classification not only organizes the vast research that has been done, but also helps to highlight connections among research areas and those which are in need of more attention.

Within this classification, the extremes of dimensionless parameters are the easiest to understand and to model. Among the future topics which we envision are more complete explorations of the internal boundaries of the diagram, where the length and time scales of swimmers and obstacles are comparable, and the interactions of swimmers are on the precipice of a transition from dilute to collective behavior.

Another future topic is further study of systems that do not fit within this three-dimensional classification space. For example, this classification can be viewed as the low Reynolds number region of a four dimensional space with Reynolds number as an additional axis. This classification also assumes that the swimmers and obstacles have a single dominant length scale and time scale. Also absent in the classification is the obstacle concentration. For small $\Bn$, this concentration will determine the flow properties of the suspending medium. For large $\Bn$, we have focused on large single (isolated) obstacles or boundaries. Depending on the system type, the obstacle concentration may be an important axis to explore. 

Nearly a century of effort combining biological, physical, and mathematical perspectives has established an incredible array of detailed knowledge about swimming in complex fluids. With our first steps to understand the simplest model systems behind us, we continue on to embrace the fantastic complexity of the biological realm. 

\vspace{.2cm}
\noindent {\it ``It takes more than a knock on the head to kill a scientist,'' said Benes. ``All that mathematics makes the skull as hard as a rock, eh?''} \cite{akk66}
\vspace{.2cm}

\begin{issues}[FUTURE ISSUES]
\begin{enumerate}
\item Additional length- and timescales provide additional axes to explore. Inertial effects would appear along another axis related to the Reynolds number. Obstacle volume fraction is another. Some systems may have multiple important swimmer length scales, like the cell body size, the flagellum diameter, and run length of a random-walk motion. Systems may have an important timescale related to their motion as individuals but also a timescale related to dynamics of the group.
\item Many studies focus on well-controlled combinations of a single type of swimmer and a single type of complex fluid, while many applications of interest are mixtures. Heterogeneous and multi-scale systems continue to need attention.
\item Within each type, the system behavior depends on the specifics of swimmer actuation, swimming gait, and deformability. This dependence will be particularly important when seeking to understand natural biological systems and engineering new synthetic systems.
\item Systems at the boundaries between identified types are expected to combine emergent features from the extremes of the classification system. Transitions between extreme behaviors may be smooth or sudden.
\item Among the eight presented system types, Type VI systems (active suspensions in fluids with bulk complex features) are particularly challenging and understudied, both experimentally and theoretically. 
\item The areas of systems and synthetic biology have expanded our understanding of internal biological processes, including how they may change in response to their environment. Coupling these internal processes with the external complex fluid behaviors described here will be an important future direction.
\end{enumerate}
\end{issues}

\section*{DISCLOSURE STATEMENT}
The authors are not aware of any affiliations, memberships, funding, or financial holdings that might be perceived as affecting the objectivity of this review. 

\section*{ACKNOWLEDGMENTS}
We gratefully acknowledge Gwynn Elfring for helpful conversations and for critically reading the manuscript. SES acknowledges the support of the NSF/NIH (DMS-1661900, DMR-2003819). PTU acknowledges the support of the NSF (CBET-0954445, DMS-1211665).

\bibliographystyle{ar-style4}
\bibliography{SwimBib}

\end{document}